\documentclass[twocolumn]{aastex63}
\usepackage{ulem}

\shorttitle{The Hydra cluster}
\shortauthors{Lonoce et al.}

\begin{document}

\title{The initial mass function and other stellar properties across the core of the Hydra I cluster\footnote{This paper includes data gathered with the 6.5 meter 
Magellan Telescopes located at Las Campanas Observatory, Chile.}}

\correspondingauthor{Ilaria Lonoce}
\email{ilonoce@uchicago.edu}

\author[0000-0001-8421-1005]{Ilaria Lonoce}
\affiliation{Department of Astronomy $\&$ Astrophysics, The University of Chicago, 5640 South Ellis Avenue, Chicago, IL 60637, USA}

\author[0000-0003-3431-9135]{W. L. Freedman}
\affiliation{Department of Astronomy $\&$ Astrophysics, The University of Chicago, 5640 South Ellis Avenue, Chicago, IL 60637, USA}

\author[0000-0002-0160-7221]{A. Feldmeier-Krause}
\affiliation{Max-Planck-Institut f\"ur Astronomie, K\"onigstuhl 17, 69117, Heidelberg, Germany}



\begin{abstract}

The Hydra I cluster offers an excellent opportunity to study and compare the relic old stellar populations in the core of its two brightest galaxies. In addition, the differing kinematics of the two galaxies allows a test of the local validity of general scaling relations. In this work we present a direct comparison employing full spectral fitting of new high-quality long-slit optical and NIR spectroscopic  data. We retrieve age, metallicity and 19 elemental abundances out to $\sim12$ kpc within each galaxy, as well as the IMF in their central regions.
Our results suggest that the inner $\sim5$ kpc region of both galaxies, despite their different masses, formed at the same time and evolved with a similar star formation time-scale and chemical enrichment, confirming their early formation in the cluster build up. Only the overall metallicity and IMF radial profiles show differences connected with their different velocity dispersion profiles. The radial trend of the IMF positively correlates with both [Z/H] and $\sigma$. While the trends of the IMF with metallicity agree with a global trend for both galaxies, the trends with the velocity dispersion exhibit differences. The outer regions show signs of mixed stellar populations with large differences in chemical content compared to the centers, but with similar old ages.

\end{abstract}



\section{Introduction} 
\label{sec:intro}

Despite its apparent simplicity, reconstructing the formation and evolution of massive elliptical galaxies is still a great challenge, and both theoretical and observational efforts are still ongoing with the aim of creating a complete assembly picture for these stellar systems. 

Large local galaxy surveys have allowed the characterization of the stellar population properties of the overall population of ellipticals, and the construction of scaling relations to derive information on their past histories, with the robustness of statistical samples (e.g.: \citealt{thomas10,sanchez12,ma14,mcdermid15}). 
More recently, it has also been possible to derive the global trend of stellar properties, including elemental abundances, as a function of the galaxy radius  (e.g.: \citealt{parikh19,zibetti20}). These studies confirmed the presence of radial gradients for a large population of local galaxies. A radial variation has been also widely observed for the stellar Initial Mass Function (IMF),  leading to the conclusion that the IMF is non-universal, among and within galaxies (e.g.: \citealt{treu10,cappellari12,conroy12,martin15b}). 

Observations of radial gradients generally support the scenario of a two-phase process for the build-up of massive ellipticals (\citealt{naab09,oser10}), with the $\textit{in situ}$ stars formed at high-redshift ($z>3$) as a consequence of a rapid cold accretion of gas \citep{dekel09}, and the $\textit{ex situ}$ ones accreted in the outskirts during a prolonged following phase.

Complementary to these global studies, generally obtained from stacked spectra, studies of a single or a few peculiar objects can offer the advantage of having higher signal to noise data, which can be studied in greater detail. This is the case for the Hydra I cluster, the object of this work, whose brightest cluster galaxy (BCG) NGC3311 has been intensively investigated together with its surrounding stellar halo and cluster companion stellar systems. From previous studies we have learnt that NGC3311, a cD galaxy with a low central surface brightness and extended radial profile, has a $\sim3$ kpc inner core characterized by an old age, super-solar metallicity and Mg and Na enhanced abundances with respect to solar values. The core is likely a relic of the \textit{in situ} stars that formed early in the first phase of galaxy formation (e.g.: \citealt{barbosa16,barbosa21a}), according to the framework of the two-phase scenario. In its very center, an irregular dust disk embeds new star formation, confirmed by the presence of bright blue spots and strong emission lines \citep{richtler20}. 
At larger radii, kinematic signatures and gradual variation and scatter in its stellar properties indicate the presence of a more complex stellar content, added in subsequent phases of accretion of material from other surrounding stellar systems (e.g.: \citealt{coccato11,ventimiglia11,arnaboldi12,barbosa16}). 
Beyond $\sim6-7$ kpc, where the contribution of dark matter increases and dominates \citep{richtler11}, a dynamically hot stellar halo extends out to $\gtrsim40$ kpc from the central BCG \citep{barbosa18}. First estimates point to a still old stellar population, similarly $\alpha$-enhanced, but much more metal poor \citep{coccato11}, although a large scatter dominates the measurements \citep{barbosa16}.
Indeed, tidal streams, dwarf galaxies (e.g. HCC 026, HCC 007) and a large number of globular clusters populate and are currently falling into the cluster core \citep{arnaboldi12}, enriching it with stars having possibly different origins. 

The brightest companion of NGC3311 is NGC3309, a massive elliptical galaxy with a line-of-sight velocity offset of $\sim250$ km/s from the BCG, and with a separation on the sky of only $100\arcsec$, but with no signs of interaction with the BCG \citep{arnaboldi12}.
Although close to each other in the cluster core, the two giant ellipticals have  different surface brightness profiles, with NGC3309 characterized by a typical R$^{1/4}$ profile, and NGC3311 by multiple components \citep{arnaboldi12}. Also their velocity dispersion profiles are remarkable different: while the central $5$ kpc region of  NGC3309 has a  symmetric negative gradient, typical of ellipticals, NGC3311 has  a peculiar inverse positive gradient. This is an important sign that suggests that the two objects have had a different formation and evolutionary history, opening the question of what has been the driver of such differences. Moreover, this peculiarity in the velocity dispersion profile offers the rare opportunity to test the local validity of the widely observed scaling relations (e.g.: \citealt{thomas10,conroy14, parikh19}).

The IMF is an essential ingredient for understanding the mechanisms of formation and evolution of stellar systems (e.g.: \citealt{conroy12,martin15b,labarbera17,vaughan18,sarzi18}). Correlations of the IMF with other stellar properties are a first tool to investigate the drivers of the IMF shape during the galaxy formation process (e.g.: \citealt{martin15c,vandokkum17,barbosa21b}). As a consequence, directly comparing the IMF radial profile of the two galaxies in this study, as well as in relation to other stellar properties, can provide insight into their formation and evolution.   

However, as discussed in our previous papers (\citealt{feldmeier20,lonoce21}), the measurement of the IMF is technically very challenging, and different assumptions, as well as the choice of method or models used, can lead significantly different results. Very high S/N ratio data are needed ($>100$\AA$^{-1}$), and a full and solid characterization of the chemical content is crucial to avoid biased results \citep{lonoce21}.

In this work we further investigate the stellar population radial profile of the two main elliptical galaxies of the Hydra I cluster, NGC3311 and NGC3309, adding for the first time details on the chemical content with the retrieval of many elemental abundances necessary to retrieve their IMF radial profile with good precision. We derive relations among stellar properties, retrieved with the same data set, analysis and models, in order to isolate possible driver(s) of the first phase of formation of NGC3311 and NGC3309. 
We also characterize part of the surrounding stellar halo, giving for the first time estimates of many halo elemental abundances. Since different chemical elements are enriched on different time-scales, we are also able to compare the star formation time-scale of the halo with respect to the central regions and find clues of their past origin.

The paper is organized as follows: in Section \ref{sec:data} we present the observations, details on the data reduction process and a focus on the observed emission lines; in Section \ref{sec:analysis} we provide details on the analysis setup, including a description on how we deal with the outer regions and the determination of systematic errors; we comment on our results in Section \ref{sec:results}, and we discuss them more broadly in Section \ref{sec:discussion}; finally we summarize the findings of this work in Section \ref{sec:conclusions}.


\section{Spectroscopic data}
\label{sec:data}

\subsection{Observations}
\label{sec:observations}
The two targets, NGC3309 and NGC3311, were observed simultaneously during the nights of April 28-29, 2019 with the Inamori-Magellan Areal Camera $\&$ Spectrograph 
(IMACS, \citealt{dressler06}) on the Baade Magellan Telescope at the Las Campanas Observatory, Chile. Indeed, their apparent proximity in the sky, i.e. only $100$\arcsec, 
and the length of IMACS longslit, $15$\arcmin, allowed us to obtain spectroscopic data out to two effective radii for both galaxies, as well as part of the 
surrounding stellar halo in both directions as shown in Figure \ref{fig:imaging}. The position angle was $\sim110^{\circ}$ East of North.
Due to the decreasing S/N of the stellar halo light at larger distances from the cluster center, as shown in Fig. \ref{fig:SN}, 
we focused the analysis only on the region between $\sim100$\arcsec\space west of NGC3309 and $\sim100$\arcsec\space east of NGC3311, and used the remaining outermost regions to evaluate the 
background and foreground, as discussed below.
We observed the sources with two grating configurations as listed in Table \ref{tab:info-obs}: with the $600-8.6$ 
grating and grating angle (GA) of $9.71^{\circ}$ to cover the optical region (i.e. $3500-6700$\AA), and with the $600-13.0$ grating and GA of $17.11^{\circ}$ to cover the 
near-IR ($7500-10500$\AA). Since the CCD consists of $4+4$ separated chips, the wavelength region of each GA configuration is divided in four sub-regions and, as a 
consequence, it has three chip gaps of $\sim50$\AA, at around $4300$, $5100$ and $5900$\AA\space in the optical. 
We made the choice of GA values in order to make sure that all the relevant spectral features did not fall on the chip gaps. From the red GA, we used only the
chip around the Calcium Triplet (CaT) feature, i.e. from $8100$\AA\space to $8900$\AA.
We observed both GA configurations for $2.33$ hours divided in seven frames of $1200$s each, and with an average seeing of $0.6\arcsec$. 
The choice of the $2.5$\arcsec\space slit width provided us with a $\lambda$-constant spectral resolution of about $5.5$\AA. We observed a small variation of the spectral resolution in the vertical direction along the slit, with values differing by $\sim10$\% from top to bottom. 

\begin{table*}
\begin{centering}
 \caption{Main features of the Hydra I spectroscopic data: grating angle (GA), date of observations, grating, slit width, position angle, obtained spectral range and 
 total exposure time.}
 \label{tab:info-obs}
 \begin{tabular}{lcccccc}
 \hline
 \hline
 Grating Angle  & Period & Grating & Slit Width  & Position angle  & Spectral Range & Exposure Time \\
    ($\degr$)   &        &         &    (\arcsec) & ($\degr$)          & (\AA)          & (min)     \\
 \hline
 9.711     & April 28-29, 2019 & 600-8.6  & 2.5 & 110 & 3500-6700  & 140  \\
 17.11     & April 28-29, 2019 & 600-13.0 & 2.5 & 110 & 7500-10500 & 140 \\
 \hline
 \end{tabular}
\end{centering}
\end{table*}

\subsection{Data reduction}
\label{sec:reduction}
The data reduction was carried out with the standard tools of IRAF \citep{tody93} and custom IDL scripts. For both GA configurations we removed cosmic rays and 
bad pixels, performed bias subtraction, flat-fielding and applied the wavelength calibration in air wavelength. The two targets were located in the lower row of four chips and 
their light, plus the light of the outer halo, plus the presence of an unforeseen foreground emission (see below and Appendix \ref{app:foreground}), entirely covered the spatial vertical direction of the 
frames, preventing us from estimating the background. 
Similarly, the upper row of chips could not be used to extract the background, again due to  the presence of stellar halo light and foreground emission. 
Therefore, we made use of the publicly available ESO tool SkyCal Sky Model Calculator (Cerro Paranal Sky Model, \citealt{noll12,jones13}). 
Given the details of the observations, i.e. telescope coordinates and altitude, time of observations and target coordinates, this tool retrieves the predicted radiance 
sky spectrum for the desired wavelength range, wavelength grid and spectral resolution. In particular, the radiance spectrum includes the following components: 
scattered moonlight and starlight, zodiacal light, molecular emission of lower and upper atmosphere and airglow continuum. 
For each reduced scientific frame, we calculated three modelled sky spectra with three different spectral resolutions, i.e. those measured at the top - middle - bottom rows of the frame, to take care of the varying spectral resolution in the slit direction. Interpolating the three spectra along the vertical direction, we computed the 2D sky model spectrum with an optimized spectral resolution. 
We then calculated, for each frame, the multiplicative factor that minimizes the difference between the tabulated and observed sky counts. 
The sky-subtracted frame was obtained by simply subtracting the modelled 2D sky from the original frame. In the optical region sky residuals were minimal. 
Only about one to two pixels per emission line still have residuals due to the unavoidable difference in the shape of the tabulated sky (boxcar) emission lines with respect to the real ones. 
All residuals have been flagged and not included in the analysis fit. However, in the red region, the subtraction of the SkyCal sky spectrum left larger residuals. 
This was due to the presence of numerous sky emission lines that scale with other sky features in non-linear ways, and the optimized solution did not always remove the presence
of residuals like in the NaI and CaT spectral features. This was more frequent in the low S/N spectra of the outer halo. 
As before, we masked the sky residual regions in the analysis.

Red GA data, obtained since 2018, have a fringing pattern all over the 2D frames. As described in \citet{lonoce21}, this effect can be efficaciously removed with the help 
of spectroscopic flat field frames taken just before and after the scientific exposures. Each scientific frame is then corrected for fringing using the flat field 
frame interpolated to its proper time. This minimizes the effect of time variability of the fringing pattern. 

\begin{figure*}[ht!]
\begin{centering}
\includegraphics[width=17.cm]{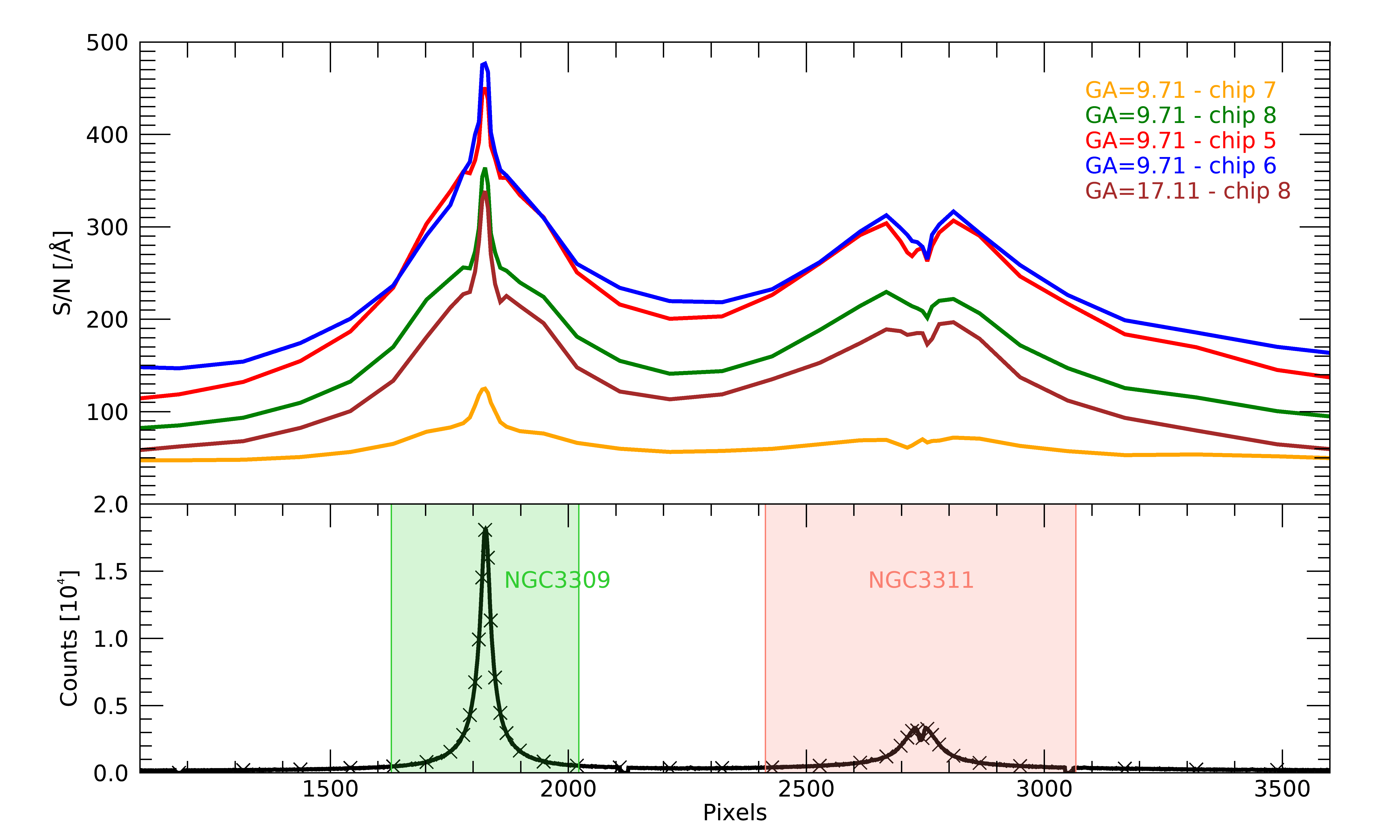}
\caption{\small{\textit{Upper panel:} S/N per \AA\space trends along the slit direction as measured for each CCD chip on the final reduced spectra. The pixel scale is $0.111$\arcsec\space per pixel. In detail: chip7-GA$=9.71^{\circ}$ 
covers $\sim3500-4300$\AA\space (orange), chip8-GA$=9.71^{\circ}$ $\sim4300-5100$\AA\space (green), chip5-GA$=9.71^{\circ}$ $5100-5900$\AA\space (red), chip6-GA$=9.71^{\circ}$ $5900-6700$\AA\space (blue) 
and chip8-GA$=17.11^{\circ}$ $\sim8000-8800$\AA\space (brown). \textit{Lower panel:} light profile along the slit that connects the two target galaxies NGC3309 (green) and 
NGC3311 (pink). Shades highlight the region within $1$R$_e$ for each galaxy, i.e. R$_e^{3309}=21.9$\arcsec\space and R$_e^{3311}=36.2$\arcsec\space \citep{arnaboldi12}. 
Crosses indicate the central position of each region where a spectrum has been extracted. }}
\label{fig:SN}
\end{centering}
\end{figure*}

The extraction of the 1D spectra was performed by means of a custom IDL code on the sum of all of the seven $1200$s reduced frames. After retrieving the curvature of 
each CCD chip in the spatial direction by fitting the position of the peak of NGC3309 along the dispersion direction, the code sums the flux in all of the chosen physical regions from 
the west stellar halo, NGC3309, the halo between the two galaxies, NGC3311 and the east stellar halo. 
These regions have been defined to ensure that the S/N in the region around $5000$\AA\space was $\gtrsim100$ \AA$^{-1}$, as shown in Fig. \ref{fig:SN}. However, as discussed in Section \ref{sec:analysis}, due to the increasing broadening of features that prevent a reliable retrieving of stellar population properties, we only used $40$ of them, from $\sim50$\arcsec\space west of NGC3309 and  $\sim70$\arcsec\space east of NGC3311.

The extracted 1D spectra have been flux calibrated by means of standard stars observed at the same airmass soon after each scientific exposure. Finally, they have been further 
corrected for tellurics $\gtrsim5000$\AA, with the software MOLECFIT \citep{molecfit1,molecfit2}.

The data reduction process was performed on each of the four chips separately. The step of the wavelength calibration relies on arc frames 
whose number of strong emission lines differ from chip to chip. This means that the quality of the $\lambda$ calibration is different depending on which chip is considered. 
As a consequence, possible $\lambda$ shifts can occur between adjacent chips. In order to attach together the $4+1$ (optical+NIR) spectral regions, we thus first retrieved 
their kinematic properties, in particular their radial velocities, which gave us an 
estimate of the relative wavelength shifts among chips. We derived the kinematics of all the extracted spectra using PPXF \citep{cappellari17} with the MILES models 
\citep{vazdekis10} and a Chabrier IMF. We found differences among radial velocities measured from different chips of the order of $100$ km/s. Spectra from GA$9.71$-chip6 and GA$17.11$-chip8, which have arcs with the higher number of lines ($\sim20$) are close to the expected values observed in the literature (see Figure \ref{fig:kinematics}). 
To homogenize the wavelengths of different chips to a common grid, we shifted all wavelengths such that the center of NGC3309 has the literature 
value of $4089$ km/s  \citep{smith04}. In this way, while homogenizing within the same spectrum along the whole wavelength range, we kept the relative velocity 
shifts among spectra from different physical regions. For a final check, we ran PPXF again on the final spectra with all of the chips combined. We found consistent 
results with literature values not only for NGC3309, as expected, but also for NGC3311 \citep{richtler11,barbosa18}. \\

\begin{figure*}[ht!]
\begin{centering}
\includegraphics[width=20.0cm]{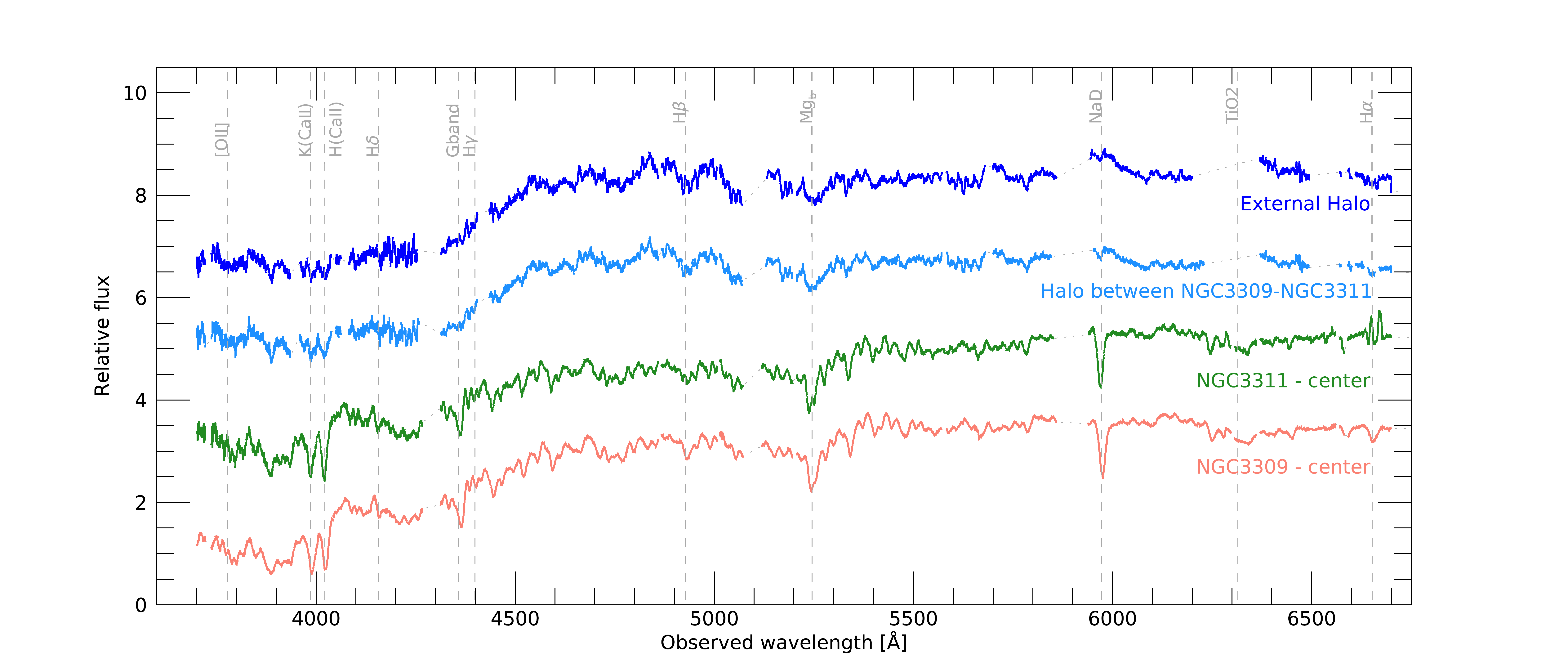}
\caption{\small{Four examples of the final reduced spectra in the optical region: the center of NGC3309 (pink), the center of NGC3311 (green), a region of the halo 
between the two galaxies (light blue), and an external region at the east of NGC3311 (blue). Some representative spectral features are shown (gray). H$\alpha$ is visible in absorption for NGC3309 and in absorption+emission 
for NGC3311 since its core has a dust disk where star formation is ongoing \citep{arnaboldi12, richtler20}. Gaps between CCD chips can be noticed around $4300$, $5100$ and $5900$\AA. Other masked regions are due to sky residuals and foreground emission (see text). Fluxes are normalized around $4500-5500$\AA\space and shifted for clarity.}}
\label{fig:spectra}
\end{centering}
\end{figure*}

Four examples of the final reduced spectra are shown in Fig. \ref{fig:spectra}: the center of NGC3309 (pink), the center of NGC3311 (green), one of the regions between the two galaxies where the stellar halo is dominant (light blue), and one from the external halo to the east of NGC3311 (blue). We intentionally omit a description of the absolute flux matching between adjacent spectral chunks since the extraction of information on stellar parameters has been carried out on each normalized chip spectrum in a parallel way, and by comparing \textit{relative} flux features with those of models, as detailed below in Section \ref{sec:alfsetup}.

\begin{figure*}[ht!]
\begin{centering}
\includegraphics[width=14.0cm]{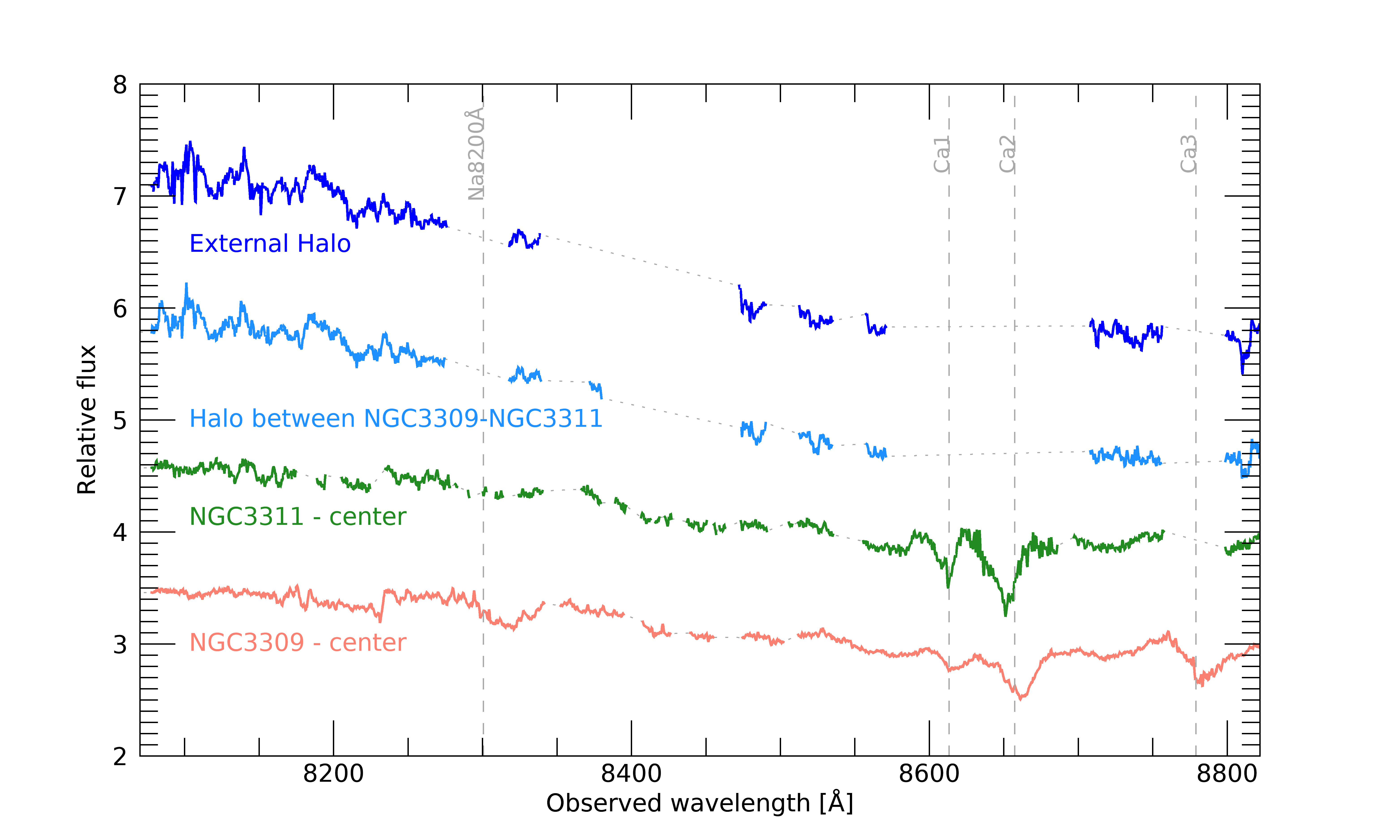}
\caption{\small{Same as Fig. \ref{fig:spectra} but in the red region around the CaT feature, as highlighted in green. All sky residuals have been properly masked and not included in the fit as described in Section \ref{sec:analysis}. Fluxes are normalized and shifted for clarity.}}
\label{fig:spectraRED}
\end{centering}
\end{figure*}

\subsection{Star formation and foreground emission}
\label{sec:starformation}
\begin{figure*}[ht!]
\begin{centering}
\includegraphics[width=16.7cm]{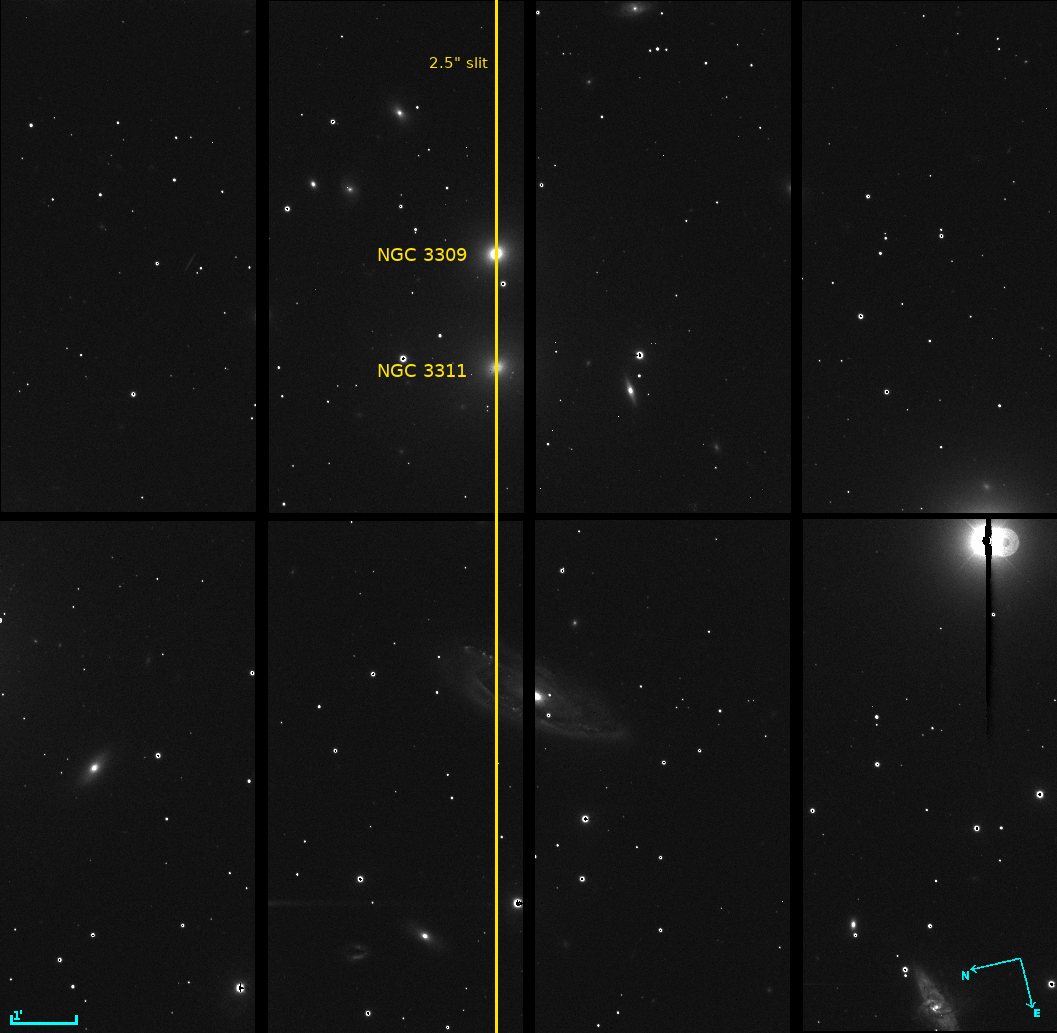}
\includegraphics[width=16.7cm]{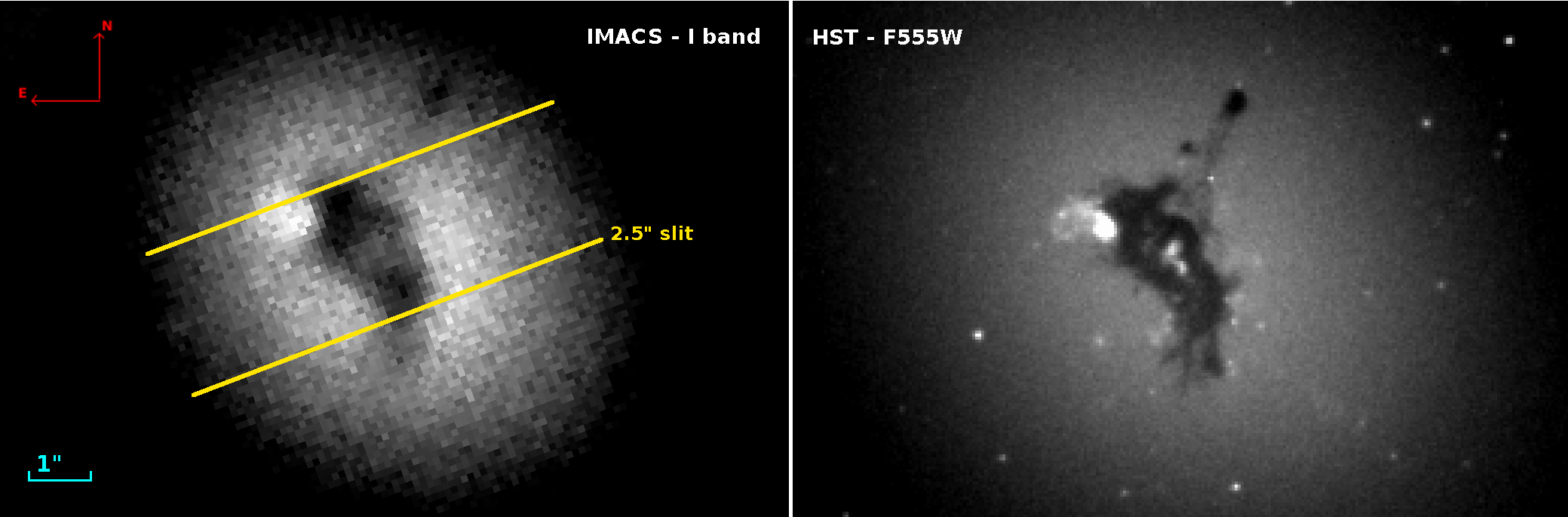}
\caption{\small{\textit{Upper panel:} B band image of the Hydra I cluster IMACS field. The vertical yellow line traces the position of the $2.5$\arcsec-wide slit used for the acquisition of our spectroscopic data. \textit{Lower panel:} NGC3311 center imaging. \textit{Left:} IMACS I band imaging; yellow lines trace the position and orientation of the $2.5$\arcsec\space longslit. \textit{Right:} HST-WFPC2 F555W imaging of the same region. Both images show the dusty central region of NGC3311, where bright spots indicate the presence of internal ongoing star formation \citep{richtler20}. 
 As a result of the orientation of the slit across this region, we could extract 1D spectra isolating the region with the brighter spot and no dust, and the dusty region with 
 the smaller bright spot, confirming the presence of a very young stellar population and detailing its chemical characteristics (see Section \ref{sec:results}).}}
\label{fig:imaging}
\end{centering}
\end{figure*}

Some visible emission lines can be observed in the central regions of NGC3311 (e.g.: H$\beta$, H$\alpha$ and [NII]) where some star formation is ongoing \citep{arnaboldi12, richtler20}. Fig. \ref{fig:imaging} shows the image of the dusty center of NGC3311 as observed by IMACS (left) in the I band, and by HST-WFPC2 
(right) in F555W \footnote{\citet{richtler20}, based on observations made with the NASA/ESA Hubble Space Telescope, and obtained from the Hubble Legacy Archive, which is a collaboration 
between the Space Telescope Science Institute (STScI/NASA), the Space Telescope European Coordinating Facility (ST-ECF/ESA) and the Canadian Astronomy Data Centre (CADC/NRC/CSA)}. 
In the left panel we highlight in yellow the orientation of the $2.5$\arcsec\space 
longslit across NGC3311. 
As observed by \citet{richtler20}, an excess of blue light is present, corresponding to the
bright spots embedded in the dust. In our IMACS data, and as better delineated by the higher resolution HST image, we can clearly identify these regions, as shown in Fig. \ref{fig:imaging}.
We then extracted the 1D spectra by keeping these regions separated, and we observed the most intense emission lines corresponding to the bright spot in the north-east 
corner of the dust structure. The spectrum extracted from the central dusty region shows moderate emission lines due to the presence of the smaller central bright spot, confirming that some star formation is ongoing in the disk. As described in Section \ref{sec:analysis}, in our analysis we fit the spectra with the inclusion of 
emission lines, and in the particular cases of the spectra around the NGC3311 center, we allowed for the presence of two stellar components, to take into account the small contribution of young stars.    

\vspace{0.5cm}

Looking closely at all extracted spectra, both around the two galaxies and along the halo, we noticed the presence of a set of foreground emission lines, including: 
[OII$3727$\AA], H$\beta$, [OIII$5007$\AA], [NI$5200$\AA], H$\alpha$ and [NII$6585$\AA] 
(see Figure \ref{fig:foregroundemission}). As fully detailed in Appendix \ref{app:foreground}, this local emission, constant along the entire observed field of view, is typical of the Warm Ionized Medium (WIM), as confirmed also by the Wisconsin H-Alpha Mapper (WHAM, \citealt{haffner03}) Sky Survey,
that observed  diffuse H$\alpha$ emission in the foreground of the Hydra I cluster. Since this emission is local and does not affect the Hydra I cluster region located 
at $z\sim0.013$, we masked all of the local emission lines in all spectra so as to not include them in the analysis.

\section{Analysis}
\label{sec:analysis}

For the analysis of the $40$ spectra extracted across the Hydra I cluster center, we adopted the publicly available full spectral-fitting code ALF 
\citep{conroy18}. The full spectral-fitting technique is ideal when dealing with large wavelength ranges and when the goal is the 
retrieval of many stellar population parameters, as demonstrated in our previous works \citep{feldmeier20,lonoce21}, and also in the literature (e.g.: \citealt{conroy12,vaughan18,barbosa21a}). Indeed, the fit is performed not only over the main 
spectral features, but over the entire wavelength range with all of the good pixels of the spectrum contributing to the fit. This allows us to exploit every part of the spectrum that 
can have a dependence on one or more parameters, thus giving  more accurate results. ALF is, in particular, optimized to fit the absorption lines of optical and NIR spectra 
of stellar systems older than $1$ Gyr. The fitting is performed with the Monte-Carlo Markov-Chain (MCMC) sampler EMCEE \citep{foreman-mackey13}.
To run ALF, we adopted the \citet{conroy18} stellar population models, with ages ranging from $1$ to $13.5$ Gyr, metallicity from $-1.5$ to $+0.2$ dex and with a large 
range of IMF slope values\footnote{The IMF slope X is defined as: $dN/dm \propto m^{-x}$, see Section \ref{sec:alfsetup}}, from $0.5$ to $3.5$, and with the possibility of using different parametrizations, as for example with one or two slopes. The spectral 
resolution is $100$ km/s over the whole spectral range of $3500-9000$\AA, which perfectly covers our data.
These models adopt the MIST isochrones \citep{choi16} and are based on the optical and NIR empirical stellar spectra presented in \citet{sanchez06} 
and \citet{villaume17}.

In order to retrieve non-solar values of many elemental abundances, we made use of the theoretical response functions of \citet{conroy18}, 
provided for a wide range of age and metallicity and for a fixed Kroupa IMF \citep{kroupa}, at the same spectral resolution of the models. 
With the help of these response functions, we were able to retrieve the following $19$ elemental abundances: 
Fe, O+Ne+S (called ``a"), C, N, Na, Mg, Si, K, Ca, Ti, V, Cr, Mn, Co, Ni, Cu, Sr, Ba and Eu.

In \citet{lonoce21} we demonstrated the importance of retrieving the elemental abundances to obtain unbiased values of the IMF and other parameters. Indeed, every spectral feature that changes as a function of the IMF shape, also changes as a function of many elemental abundances. 
This is especially important for a full spectral-fitting analysis, since each pixel, with its own stellar parameter dependencies, contributes to the fit. 
To minimize the possible biases affecting the IMF, we thus chose to fit all of the available elemental abundances as free parameters.

\subsection{ALF settings}
\label{sec:alfsetup}

We prepared our spectra by transforming wavelengths to vacuum, masking gaps and bad pixels, and set up ALF with the following characteristics:
\begin{itemize} 
\item{MCMC parameters: we generally fit with a number of walkers, nwalker$=1024$, a number of steps during the burn-in phase, nburn$=10^4$, and a number of steps after 
the burn-in phase, nmcmc$=100$. In cases of insufficient convergence of any parameter (typically in the outer regions), we repeated the fit increasing nwalker as needed.}
\item{Fit type: to include all possible elemental abundances as well as the IMF as free parameters, we adopted the full mode fitting, which allows the retrieval of up to $46$ parameters including: 
all stellar population properties (21), kinematics (up to 4 components), emission lines (8), two-component star formation history (2) and non-constant IMF (up to 4 components). Additional ``nuisance'' parameters are also included to correct for stellar evolution and data uncertainties (7).}
\item{IMF parametrization: we based our main analysis adopting a single power-law IMF slope of the form $dN/dm \propto m^{-x}$, with a fixed lower cutoff of $0.08$ M$_{\odot}$. Above $1$M$_{\odot}$ the slope is fixed to $2.3$, i.e. to the Salpeter value \citep{salpeter}. We have also repeated the whole analysis with a double power-law IMF, retrieving X1 (from $0.08$ to $0.5$ M$_{\odot}$) and X2 (from $0.5$ to $1$M$_{\odot}$). However, as discussed in Section \ref{sec:discussion}, the degeneracy between X1 and X2 is high, as already noted in \citet{lonoce21} and \citet{feldmeier21}, preventing the retrieval of solid results. IMF slope values span from $0.5$ to $3.5$.}
\item{Stellar components: ALF allows a simultaneous fit to two stellar population components with different ages. The fit retrieves the age and the mass fractions of the two components. All other parameters are the same as the main component. In our analysis, we allowed the presence of a secondary 
stellar population only in the four spectra extracted from the center of NGC3311, where there are signs of ongoing star formation (see Section \ref{sec:starformation}), and in the very center of NGC3309. We highlight, however, that the minimum age allowed in ALF is $0.5$ Gyr, and thus we can only give an upper limit of the age of the younger component.}
\item{Parameter ranges: for each stellar property, a uniform prior range is set in a customized way. 
Ages run from $0.5$ to $14$ Gyr, metallicity from $-1.9$ to $0.3$ dex and the IMF slope from $0.5$ to $3.9$.
For elemental abundances we started with fixing the interval from $-0.3$ to 
$+0.5$ dex, with the exception of Na which was allowed up to $+1.0$ dex. Since our spectra span from halo regions to the centers of the two ellipticals where stellar population 
properties can be largely different, these ranges have been adapted accordingly for each spectrum. 
As a consequence, in the outer regions we allowed the parameters to reach higher values, e.g. $>\pm1$ dex, if needed. We caution that in these cases the results could suffer from systematic uncertainties due to model extrapolation (considering that response functions are provided for values $\pm0.3$ dex, with some exceptions). A special case is potassium, where its only strong feature in our wavelength range at around $4100$\AA\space could not be well fit by models even with [K/H] $>3.0$ dex. We fixed the maximum limit at $3.0$ dex and tested that this assumption does not impact the determination of all of the other stellar population parameters.}
\item{Wavelength ranges: as discussed in Section \ref{sec:data}, each final spectrum has three wavelength gaps as a result of the subdivision of the separate CCD chips. 
To avoid mismatched flux alignment between adjacent chips, we imposed the fit to be performed in the following five separated wavelength regions: $3650.3-4207.6$\AA, 
$4253.3-5001.8$\AA, $5048.9-5777.8$\AA, $5857.4-6606.7$\AA\space and $7966.3-8727.7$\AA. Within each wavelength range the spectrum and the model are continuum matched by means of a polynomial function with one order per $100$\AA.}
In the outer halo spectra we masked many pixels at long wavelengths due to strong sky residuals, as described in Section \ref{sec:data} and shown in Figure \ref{fig:spectra}.
\item{Emission lines: having masked all of the foreground emission lines (as described in Section \ref{sec:data} and in Appendix \ref{app:foreground}), we fit the 
local emission lines of the Hydra I galaxies, as allowed by ALF. The lines are: Balmer  (H$\delta$, H$\gamma$, H$\beta$ and H$\alpha$, with line ratios
assumed from Case B recombination \citealt{osterbrock89}), [OII]$3726-3729$ , [OIII] $4959-5007$, [NI]$5200$ and [NII]$6548-6583$, where all doublets have relative strengths adopted from Cloudy models \citep{cloudy}). Their retrieved intensity and kinematics are discussed in Appendix \ref{app:gas}.}
\end{itemize}

After each fit, we processed the results as suggested in the ALF documentation. This includes that the total metallicity [Z/H] and the [Fe/H] abundance have been combined by adding the two quantities together. All elemental abundances, provided by the models in relation to the total H, have been properly transformed in relation to Fe (as we show our results in Section \ref{sec:results}). 
In particular, O, Ca, Mg, Ti and Si have been corrected with the library correction factors from \citet{schiavon07} and \citet{bensby14}, as suggested by the ALF documentation. This is to compensate for the fact that models with non-solar values of elemental abundances are built with stars from the solar neighborhood. We note that these corrections are more important for lower metallicity values (e.g.: $\sim0.4$ dex at around [Z/H]$\sim-1.5$ dex). As a consequence, the outer regions and halos are mostly affected by this approximation. Moreover, excluding C, N, Cr, Ni and Na that so far have not shown the need for these corrections at low Z, all other elements have not yet been tested and corrected.

\begin{figure*}[ht!]
\begin{centering}
\includegraphics[width=20.0cm]{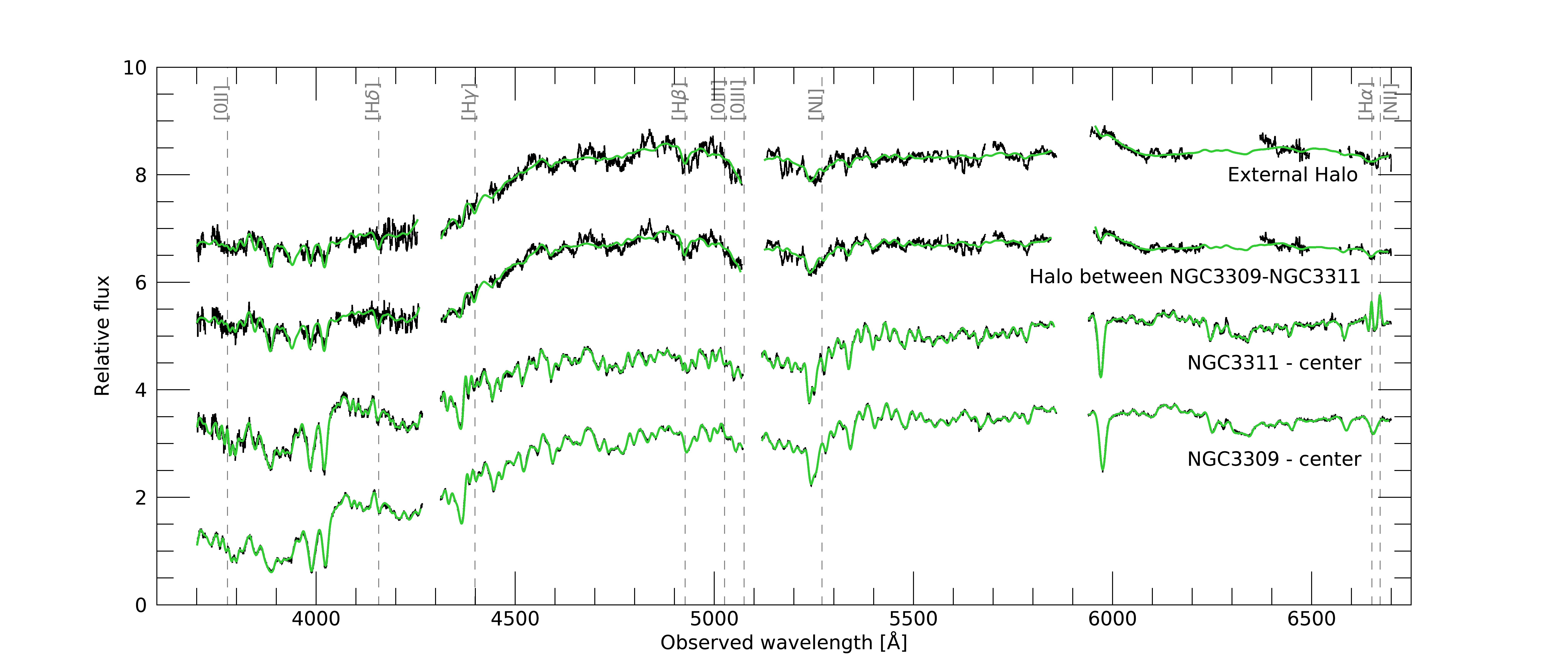}
\caption{\small{Same four example spectra of Figure \ref{fig:spectra} (center of NGC3309, center of NGC3331, halo between the two galaxies and external halo) in the optical region, and the best-fit spectra  (green). Fitted emission lines are indicated by grey vertical dashed lines. Fluxes are normalized around $4500-5500$\AA\space and shifted for clarity.}}
\label{fig:spectrafit}
\end{centering}
\end{figure*}

We have carefully checked the full convergence of each parameter by directly looking at its MCMC chain. In particular, we considered the end of the chains generally including their final $1$\% steps (i.e. $\sim1000$ steps). At the same time we verified that the values spanned by each parameter did not hit a prior limit. In cases where these conditions were not satisfied, the fits have been repeated with wider setup constraints. Figure \ref{fig:spectrafit} shows four examples of the fit obtained by ALF for the same four spectra from Figure \ref{fig:spectra}, i.e. the center of NGC3309, the center of NGC3311, one from the halo between the two galaxies and one for the external halo.

\subsection{Outer Regions}
\label{sec:outer}

The stellar halo regions, located at $>1$R$_e$ from each galaxy center (see Figure \ref{fig:SN}) and not dominated by the two galaxies' light, are characterized by having lower S/N spectra, higher velocity dispersion ($>250$ km/s) and likely a more complex stellar population composition \citep{barbosa16}. Similar characteristics also hold, in a gradual way, in the annular regions included between $\sim10\arcsec$ from the center of each galaxy (i.e. at $0.45$R$_e$ for NGC3309 and at $0.28$R$_e$ for NGC3311) and their effective radius.

The determination of the stellar parameters in these regions is more difficult, not only due to the noise and faintness of the features, but also because it is affected by possible biases due to the fact that we fit a simple stellar population model where a mix of multiple stellar populations may be the case.
We will refer to the regions from $\sim10$\arcsec\space to $1$R$_e$ as \textit{outer regions}, to distinguish them from those of the inner galaxies, which are, in contrast, well described by a single stellar population and with homogeneous stellar properties. 

Since in the outer regions we faced these above-mentioned difficulties in the retrieval of their stellar properties, we decided to fix the kinematic values in the ALF fit (described in Section \ref{sec:alfsetup}), as obtained by fitting only the spectrum from chip 5 (around $5500$\AA) with the ALF super-simple mode. We further fix age and metallicity in the halo regions with values again obtained from the super-simple mode fit. The results obtained  are consistent both with previous values and with the expectations of spectral indices.  For more details on these choices, see Appendix \ref{app:outer}.

\subsection{Systematic errors}
\label{sec:syst}

We estimated the systematic errors by considering that different regions of the spectra hold different kinds of information on the stellar population properties, and thus that fitting only a specific wavelength range can bring biased results in one or more parameters. Repeating the same fit on different wavelength ranges, therefore can provide an estimate of such uncertainties.

We analysed systematics in a slightly different way for the centers of the two galaxies and for the outer regions and halos. For the central regions, since their stellar populations exhibit similar values, we created stacked spectra to increase the S/N and highlight possible biases. In particular the stacked spectra of NGC3309 and NGC3311 are the sum of their innermost $<3\arcsec$ spectra (i.e. $\sim6$ spectra each). 
We then tested the differences when fitting the whole spectral range, without the bluest region ($<4200$\AA), without the red region ($>8000$\AA) and excluding the region $>6400$\AA. We then added in quadrature the standard deviation obtained from these fits for each stellar parameter to their statistical errors. 

For the outer regions, due to the increasing velocity dispersion and complexity of their stellar content, we preferred not to create stacked spectra but instead to consider three single spectra chosen as representative for three regions, i.e. one for the outer halos, one for the halo between the two galaxies and one for the outer region in the middle between the halo and galaxy centers. The  tests performed are the same as for the stacked spectra, and their standard deviations have been added to the statistical errors in the same way. Results in Figures \ref{fig:ageZimf} and \ref{fig:elements} show the final values with both statistical and systematic errors. 

Briefly, in the galaxy centers we found systematic error values of $\sim\pm1$ Gyr for age, $\pm0.04$ dex for metallicity, $\pm0.25$ for the IMF slope and $\pm0.08$ dex on average for the elemental abundances. For the outer regions and halos we obtained systematic errors on metallicity from $\pm0.3$ to $\pm0.5$ dex, and on elemental abundances on average from $\pm0.2$ to $\pm0.4$ dex. The IMF slope in the halos could not be constrained with our data and models as explained in Section \ref{sec:spp}. The systematic uncertainties we found in these regions on the IMF are indeed high with values around $\pm1$.


\section{Results}
\label{sec:results}
In this Section we present the results from the analysis described in the previous Section \ref{sec:analysis}, obtained when fitting 
only one IMF slope. 
A comparison with the results with two IMF slopes is detailed in Section \ref{sec:1vs2imf}.
All our results, as a function of the distance from the center of the two galaxies, are shown in Figures  \ref{fig:ageZimf}, \ref{fig:elements}, and \ref{fig:kinematics}.  

In the following Subsections we focus on the stellar population properties (Section \ref{sec:spp}) and on the kinematic results (Section \ref{sec:kinematics}). Kinematics of the gas emission component can be found in Appendix \ref{app:gas}.

\subsection{Stellar population properties}
\label{sec:spp}

\begin{figure*}[ht!]
\begin{centering}
\includegraphics[width=18.cm]{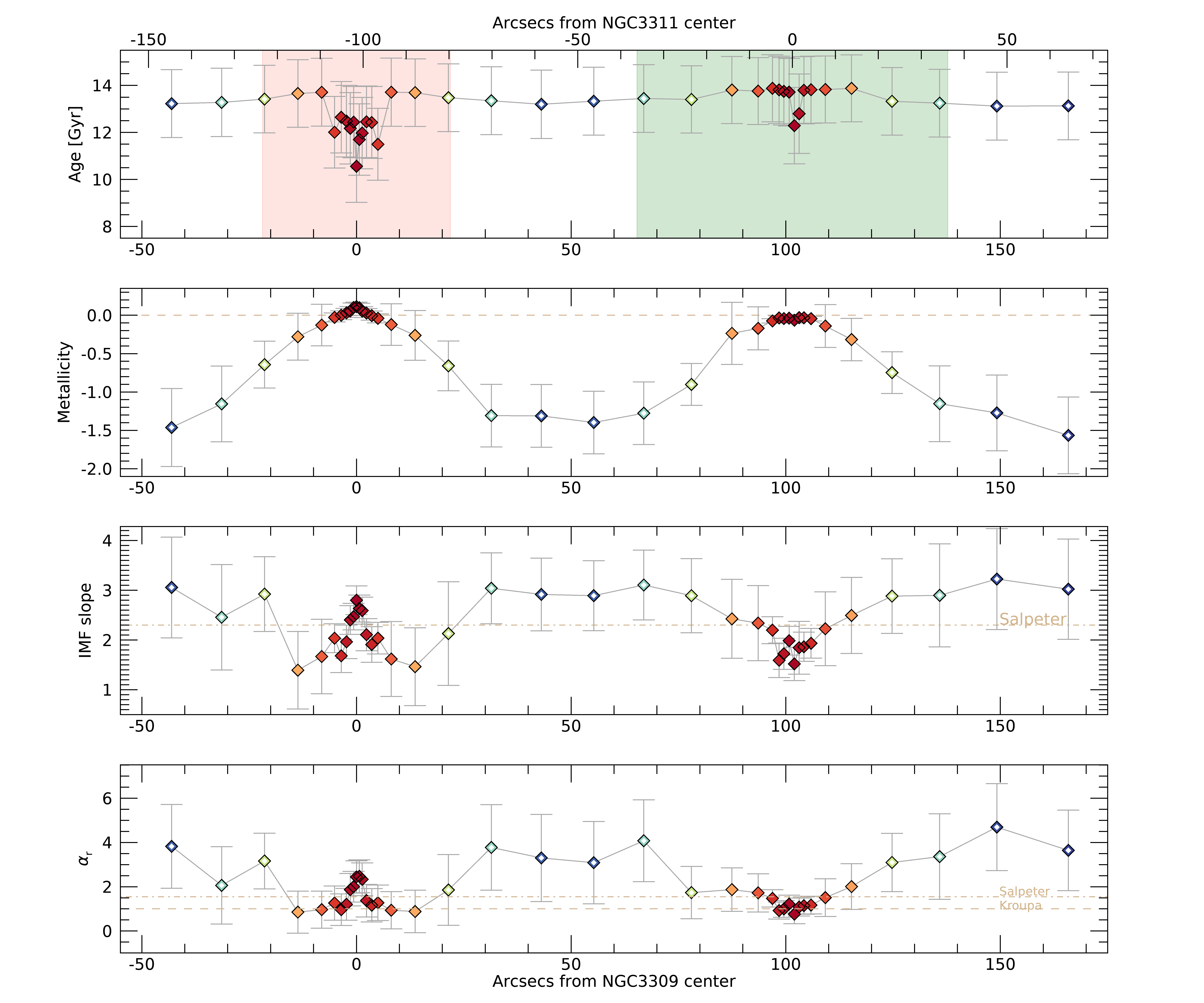}
\caption{\small{Age, metallicity and IMF slope trends across the Hydra I cluster center as retrieved with ALF. West is on the left, East is on the right. The bottom panel 
shows the derived mismatch parameter $\alpha_r$.
Dash-dotted horizontal tan lines in the IMF and $\alpha_r$ panels outline the corresponding values for a Salpeter IMF, dashed those for a Kroupa IMF. Open diamonds refer to regions where the results are less robust due to lower S/N, velocity dispersion broadening and/or lack of important features. }}
\label{fig:ageZimf}
\end{centering}
\end{figure*}

\begin{figure*}[ht!]
\begin{centering}
\includegraphics[width=18.2cm]{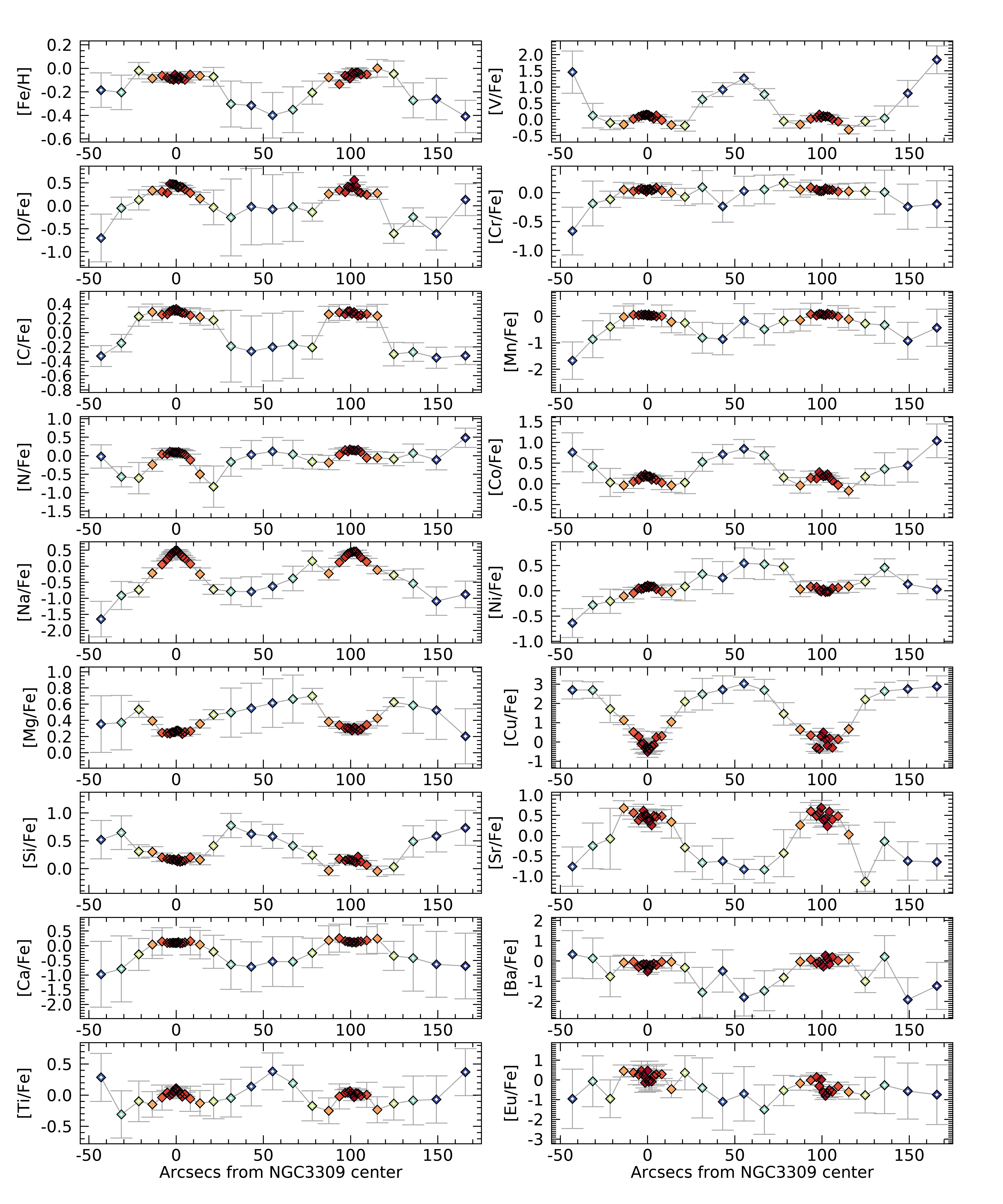}
\caption{\small{Similar to Figure \ref{fig:ageZimf} but for all the retrieved elemental abundances.}}
\label{fig:elements}
\end{centering}
\end{figure*}

Our stellar population results, obtained by fitting with ALF the $40$ spectra with the setup described in Section \ref{sec:analysis}, are shown in Figures \ref{fig:ageZimf} and 
Figure \ref{fig:elements}. The first set of plots shows the retrieved age, metallicity ([Z/H]+[Fe/H]), IMF slope and the derived mismatch parameter $\alpha_r$. $\alpha_r$ is 
defined as the ratio between the M/L in the $r$ band obtained from the best fit model, and the M/L of the same model but with a Milky-Way IMF (Kroupa): 
$(M/L)/(M/L)_{MW}$. While a value of $\alpha_r=1$ corresponds to a Kroupa IMF by definition, a value of $\alpha_r=1.55$ corresponds to a Salpeter IMF, as indicated by the horizontal lines in the two bottom panels of Figure \ref{fig:ageZimf}. Figure \ref{fig:elements} shows the results of the elemental abundances with respect to Fe. K is not shown since models could not converge even with values $>3.0$ dex (see Section \ref{sec:alfsetup}). In all plots, error bars include systematic errors as discussed in Section \ref{sec:syst}.

The results show rather constant old ages, with no visible trend from the center of the two galaxies to the external regions, with values around $13$ Gyr. This behavior is consistent with previous literature results, e.g. \citet{coccato11}, \citet{loubser12} and \citet{barbosa16}, but also in contrast with the latest findings of \citet{barbosa21a} who found a negative gradient. However, \citet{barbosa21a} show the radial results of all the Voronoi bins around NGC3311, and by inspecting their figure $5$ the sharp age gradient is mostly caused by ages as young as $5$ Gyr measured along the major axis, at a position angle PA$\sim32$\degr that is nearly orthogonal to the one we analyzed in this work (PA$\sim108$\degr); whereas the ages near PA$\sim108$\degr are higher, around $8-9$ Gyr.
NGC3311 hosts a dust disk, as discussed in Section \ref{sec:starformation}, where some level of star formation is still ongoing. As mentioned in Section \ref{sec:alfsetup},  in those spectra corresponding to the regions with dust, we fit two stellar populations to take into account the presence of the younger component. The main components have similar old ages (i.e. $\sim13$ Gyr, as shown in Figure \ref{fig:ageZimf}) as do the other surrounding central regions of NGC3311. The younger components have ages $\sim1$ Gyr with mass fractions below $1\%$. 

The stellar metallicity in the center of the two galaxies reaches solar values, while in the outer regions we obtain a negative gradient toward sub-solar values down to $\sim -1.5$ dex. Close to the center of NGC3309 the metallicity shows a clear negative gradient starting from super-solar values around $\sim0.2$ dex, while in the center of NGC3311 the metallicity trend is flat around solar values. This particular behavior is similar to the velocity dispersion trend shown in Figure \ref{fig:kinematics}, and it will be discussed in Section \ref{sec:correlations} where correlations among parameters are analyzed. 
The solar metallicity values that we found in the center of NGC3311 are in slight tension with those obtained by \citet{barbosa21a} who report a higher [Z/H]$\sim0.2$ dex. This gap could be attributed to the different adopted models. Indeed, \citet{barbosa21a} used the EMILES models \citep{vazdekis16}, which are known to have a difference of the order of $\sim0.1$ dex with the \citet{conroy18} models for old and solar/supersolar metallicities (see e.g. \citealt{feldmeier20}, \citealt{lonoce21}).
The total metallicity shown in Figure \ref{fig:ageZimf} does not show significant differences among the western, eastern and inner halos, however the [Fe/H] alone (see Figure \ref{fig:elements}) presents slightly higher values in the western region (on the left of NGC3309 in the plot). This is again in agreement with previous findings of \citet{barbosa16}.  

Interestingly, also the IMF slope trend shows some similarity with the metallicity, in particular in the regions belonging to the two galaxies, where it is better constrained. NGC3309 shows a clear negative IMF gradient, from super-Salpeter (i.e. bottom-heavy) values in its very center to a top-heavier IMF at around $10\arcsec$,
confirming the typical trend found for local ellipticals (e.g.: \citealt{martin15b,labarbera17,sarzi18}). On the contrary, the IMF profile of NGC3311 is flat in its center at sub-Salpeter values, with mild signs of a positive gradient from $\sim5\arcsec$. We stress that the IMF slope values beyond $10\arcsec$ for both galaxies are not robust as the analysis suffers from lack of IMF sensitive features, low S/N and larger velocity dispersion broadening, as reflected in their large error bars.\\
The mismatch parameter $\alpha_r$ obviously has a similar trend as the IMF slope, holding the same information. We decided to also show it in this form since, being unbounded to a particular parametrization of the IMF, it is more useful for a comparison of our results with other analyses obtained with different stellar population models. 

Elemental abundances have been precisely retrieved in the center of the two galaxies  where we found solar or super-solar values with typical errors of $0.06$ dex. [Cu/Fe], [Sr/Fe] and [Eu/Fe] have larger uncertainties also in the galaxy centers (i.e. $0.2$ dex) since the fitted wavelength ranges do not include strong features sensitive to these elements. Some elements have clear negative gradients around the centers, like [Na/Fe], [Ti/Fe], [C/Fe], [O/Fe], [V/Fe] and [Co/Fe], others have flat trends, and only [Cu/Fe] has a positive gradient. We note a very close similarity of the chemical content between the two galaxy centers, which is valid for all elements. This important result will be discussed in Section \ref{sec:discussion}.
In the halos, elemental abundance values are typically different from the inner galaxy regions, reaching in several cases extreme values beyond those provided by models and thus subject to further uncertainty due to extrapolation. For example, in the case of copper,  the extrapolation occurred up to $3.0$ dex. Generally, we do not find evident differences from values retrieved from the western, eastern or inner halos, as also confirmed by the total metallicity trend.
We also calculated the $\alpha$-elements enhancement trend by averaging together C, O, Mg, Ca, Si and Ti. The derived [$\alpha$/Fe] has a value of $\sim0.2$ dex in the center of NGC3309 and NGC3311, and a mild negative gradient toward their outskirts to around the solar value. However, the large scatter (i.e. $\sim0.25$ dex) prevents a robust confirmation of an actual gradient. Results from \citet{coccato11}, \citet{loubser12} and \citet{barbosa16} show slightly higher values ($0.3-0.4$ dex), but due to our large scatter, it is still consistent with our findings. More discussion on the [$\alpha$/Fe] trend is presented in Section \ref{sec:alphafe}.

\subsubsection{One versus two IMF slopes}
\label{sec:1vs2imf}
\begin{figure*}[ht!]
\begin{centering}
\includegraphics[width=16.5cm]{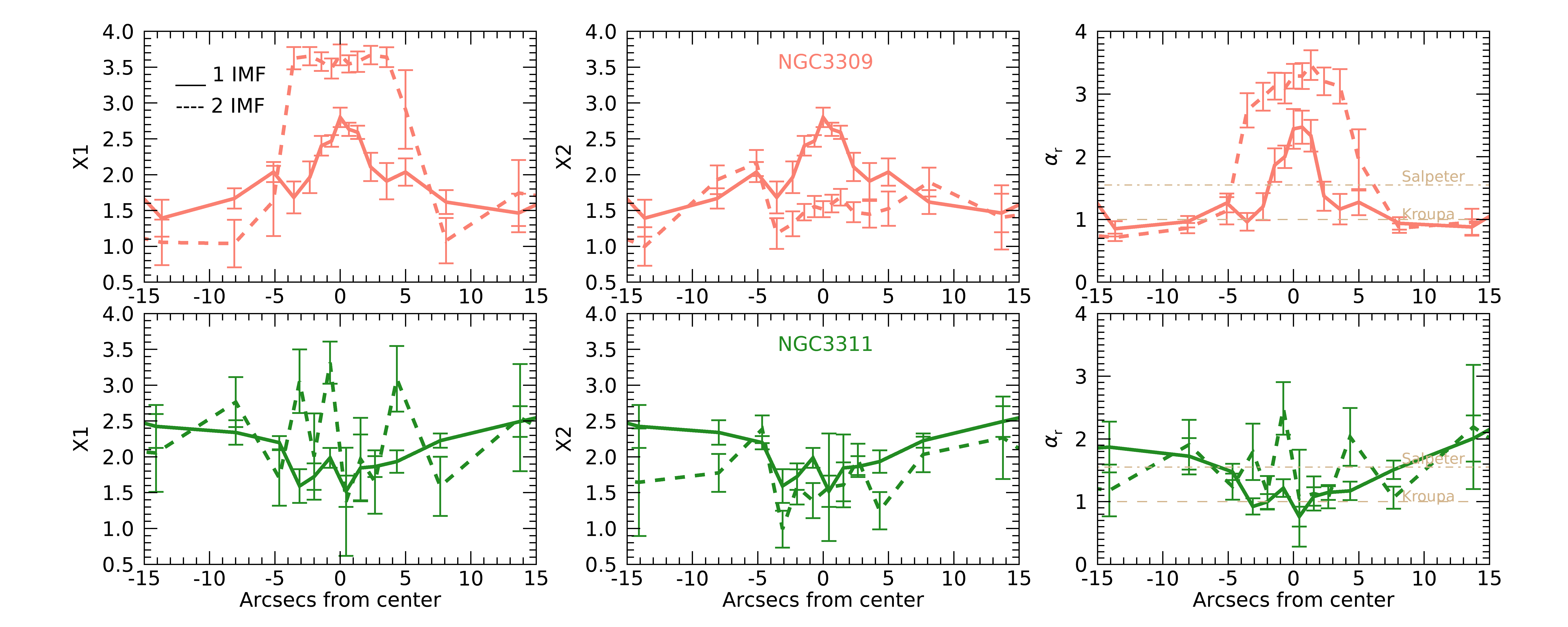}
\caption{\small{Comparison of the retrieved IMF when adopting 1 vs 2 slopes in the center of NGC3309 (upper panels) and NGC3311 (lower panels). Solid lines refer to a single slope IMF while dashed lines to a double slope IMF. The right panels show the mismatch parameter $\alpha_r$ obtained in both cases; dashed and dotted-dashed horizontal tan lines indicate the Kroupa and Salpeter value respectively. The two IMF slopes are mutually degenerate when fitted together, causing X1 to hit unnatural high values at the edge of model limits. To better highlight the uncertainties coming from the fit alone, in these plots error bars do not include systematic errors.
}}
\label{fig:1-2slope}
\end{centering}
\end{figure*}

As detailed in Section \ref{sec:analysis}, we have assumed a single slope IMF as a baseline. However, we have also performed the same fits assuming a double-slope IMF, with the first slope X1 describing the IMF in the mass range $0.08-0.5$ M$_{\odot}$, and the second one X2 in the range $0.5-1$ M$_{\odot}$. Above $1$ M$_{\odot}$, the IMF is again Salpeter. While the other parameters are not affected by this change, the retrieved IMF values are visibly different, as shown in Figure \ref{fig:1-2slope}. There, we focus on the comparison of the retrieved IMF in case of one IMF slope (solid lines) with the case of two IMF slopes (dashed lines) in the centers of NGC3309 (upper panels, pink) and NGC3311 (lower panels, green). When switching to two IMF slopes, we found very high X1 values in the center of NGC3309, reaching the limit of the allowed model values. In NGC3311 instead, we observe a larger scatter in X1, with adjacent points going from Kroupa-like to bottom-heavy IMF almost alternating. This behavior is due to the mutual degeneracy between X1 and X2, as also observed in \citet{lonoce21}. We carefully checked the cross-correlation ellipses between X1 and X2 and confirmed high levels of correlation, with a mean Spearman correlation coefficient $\rho=-0.45$ with $p=0.009$. We conclude that with our set-up, the best choice is to adopt only one IMF slope. Some level of correlation with other parameters of the fit is still present for the case of one IMF slope, e.g. age and metallicity (and slightly Na and Ti, see Section \ref{sec:correlations} and Appendix \ref{app:fitcorr}), but with lower values (Spearman coefficient $\rho\sim0.30$), well within their final error.  

\subsection{Kinematics}
\label{sec:kinematics}

\begin{figure*}[ht!]
\begin{centering}
\includegraphics[width=15cm]{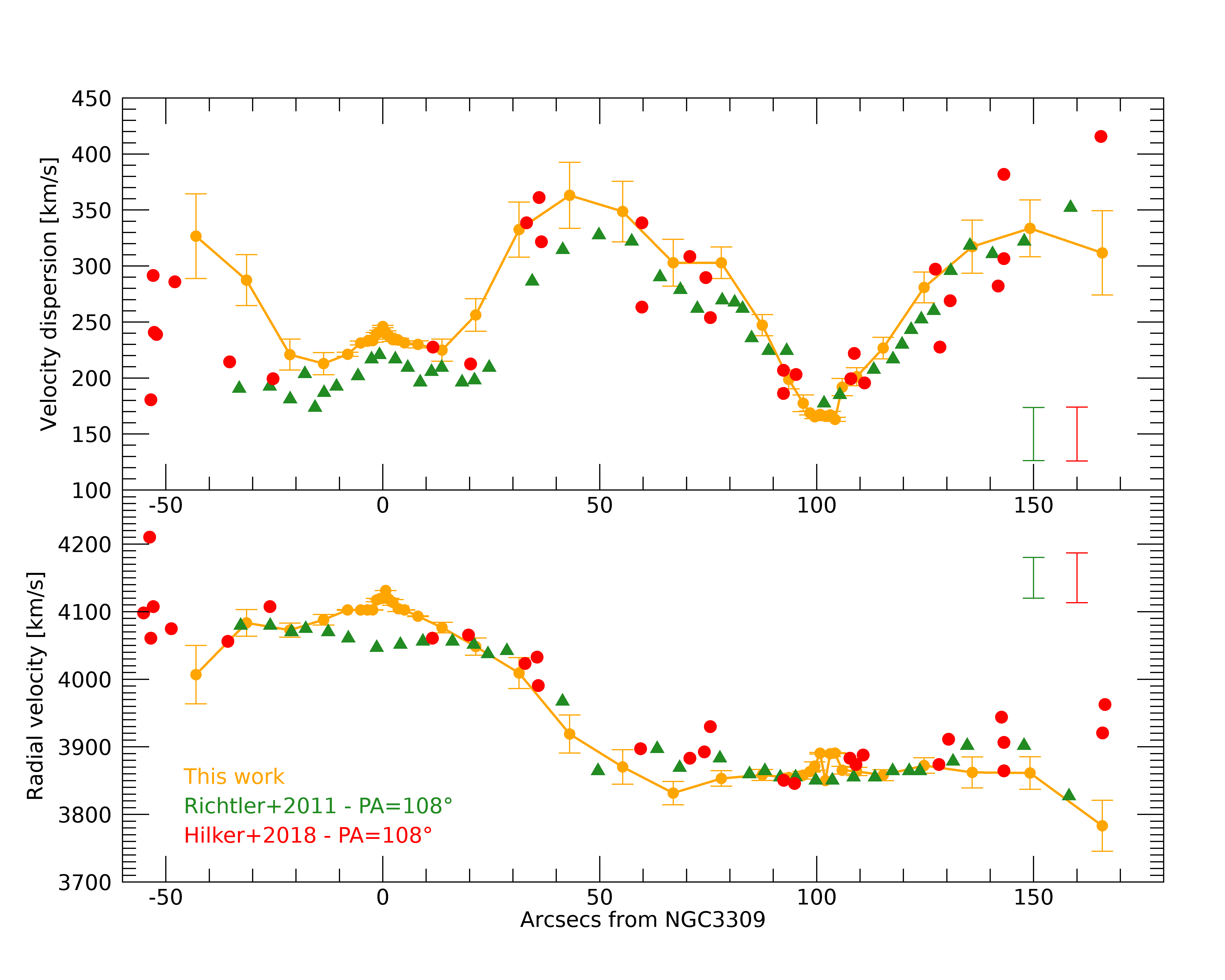}
\caption{\small{Velocity dispersion (upper panel) and radial velocity (lower panel) profiles as retrieved with ALF (orange). 
Green triangles and red circles show an indication of the trends retrieved by the works of \citet{richtler11} and \citet{hilker18}, respectively, as they appear in figure $6$ of \citet{hilker18} for the case of position angle $108$\degr.  Typical error bars are shown on the right of both panels.
}}
\label{fig:kinematics}
\end{centering}
\end{figure*}

Our kinematics results for the stellar component alone are shown in Figure \ref{fig:kinematics} as orange lines. As discussed previously, we derived the kinematics values with ALF in full mode over the entire spectra only in the centers of both galaxies, i.e. within $\sim10\arcsec$, while for outer regions and halos we relied our measurements on those extracted from fitting the chip 5 spectra (from $5100$\AA\space to $5900$\AA) with ALF in super-simple mode (Appendix \ref{app:outer}). This way we retrieved well-converged values that are in agreement with the literature. Indeed, in Figure  \ref{fig:kinematics} we have also plotted the estimated trends of the kinematic values from \citet[green triangles]{richtler11} and \citet[red circles]{hilker18}, as they appear in figure $6$ of \citet{hilker18}. They show good agreement also in regions of low S/N, particularly in the halo between the two galaxies.

The velocity dispersion profile across the core of the Hydra I cluster shows rather high values in the halo regions around $300-350$ km/s, and rapid drops in the proximity of the two galaxies, toward $\sim200$ km/s. 
In the center of NGC3309 there is a negative and symmetric gradient of $\sigma$, typical of elliptical galaxies, starting from its center toward $\sim10$\arcsec\space , with values running from $\sim250$ to around $200$ km/s.  
On the other hand, in the center of NGC3311, the $\sigma$ profile is flat at around $160$ km/s within its innermost $10$\arcsec. 
As in \citet{barbosa21b}, we will use this difference in the velocity dispersion profiles to test the validity of correlations between stellar population parameters, and in particular the IMF, with $\sigma$ (see Section \ref{sec:correlations}).  

The radial velocity profile shows a difference of $\sim250$ km/s between the two galaxies, with NGC3309 in the foreground with respect to NGC3311. As discussed in \citet{richtler11},  NGC3309 is presumably spatially at a larger distance from the BGC which resides at the center of the cluster potential. 

Similarly to its $\sigma$ profile, NGC3309 presents a (small) negative gradient in its center, not observed in \citet{richtler11} data. Instead, NGC3311 still has a flat trend. No internal rotation is visible for both objects along the adopted position angle.


\section{Discussion}
\label{sec:discussion}

With the results shown in Section \ref{sec:results}, we have provided a detailed description of the stellar population properties across the two main galaxies of the Hydra I cluster, giving for the first time an extensive picture of their chemical content and IMF.  Studying the stellar properties of these two companion galaxies, including their surrounding halos, with the same data set, as well as the same type of analysis and models, gives us the possibility of directly interpreting  observed differences between the galaxies. This is fundamental to provide unbiased constraints on their assembly history. 
In this section, we will discuss the results, focusing on both the central and outer regions, in comparison with the literature, as well as on the obtained correlations among parameters (detailed in Section \ref{sec:correlations}).

\subsection{NGC3311 in the literature}
Regarding NGC3311, the BCG of the cluster, we already have a description of its main stellar properties, for example, \citet{coccato11}, \citet{loubser12}, \citet{arnaboldi12}, \citet{barbosa16,barbosa18,barbosa21b}, including  information on the radial trend of its IMF \citep{barbosa21a}. Despite the fact that these works have been undertaken with different data sets (thus with different wavelength ranges, etc...), different kinds of analysis (spectral index fitting or full spectral fitting) and different families of models, the results on age and metallicity are 
generally in agreement, pointing to an old ($>12$ Gyr) and solar metallicity population in its center. We note an exception in \cite{barbosa21b}, who find a negative age gradient, in contrast with these other findings. 
As already mentioned in Section \ref{sec:results}, our general findings are consistent with this picture, and add more information to the observed trends of $18$ elemental abundances. In particular, the comparison of the retrieved IMF in the form of the mismatch parameter $\alpha_r$ with the results of \citet{barbosa21a} (performed with different models, thus with a different parametrization of the IMF) across the direction of our longslit, is very good within the innermost $20\arcsec$ of NGC3311, showing a flat trend around $\alpha_r\sim1.5$.

\subsection{NGC3309 and NGC3311 centers}
\label{sec:centers}

Focusing on the central regions (within $20$\arcsec) of both NGC3309 and NGC3311, where we have retrieved all the stellar parameters with high precision, we found a general agreement with similar available values found in the literature both in the center (e.g.: \citealt{graves08,johansson12,worthey14,conroy14,gu22}) and up to $\sim1$R$_e$ (e.g.: \citealt{feldmeier21,parikh19,newman17,vandokkum17}) of local elliptical galaxies, showing in some cases also similar trends with R. Nevertheless, there are few studies for which as many elemental abundances for single objects can be retrieved; to date these have generally focused on the center of galaxies and with stacked spectra. Only recently have stellar population synthesis models with non-solar abundances and very high quality spectroscopic data become available. In the near future we will be able to compare these results in detail with much larger samples both in the central and outer regions of nearby galaxies. 

\begin{figure*}[ht!]
\begin{centering}
\includegraphics[width=18.5cm]{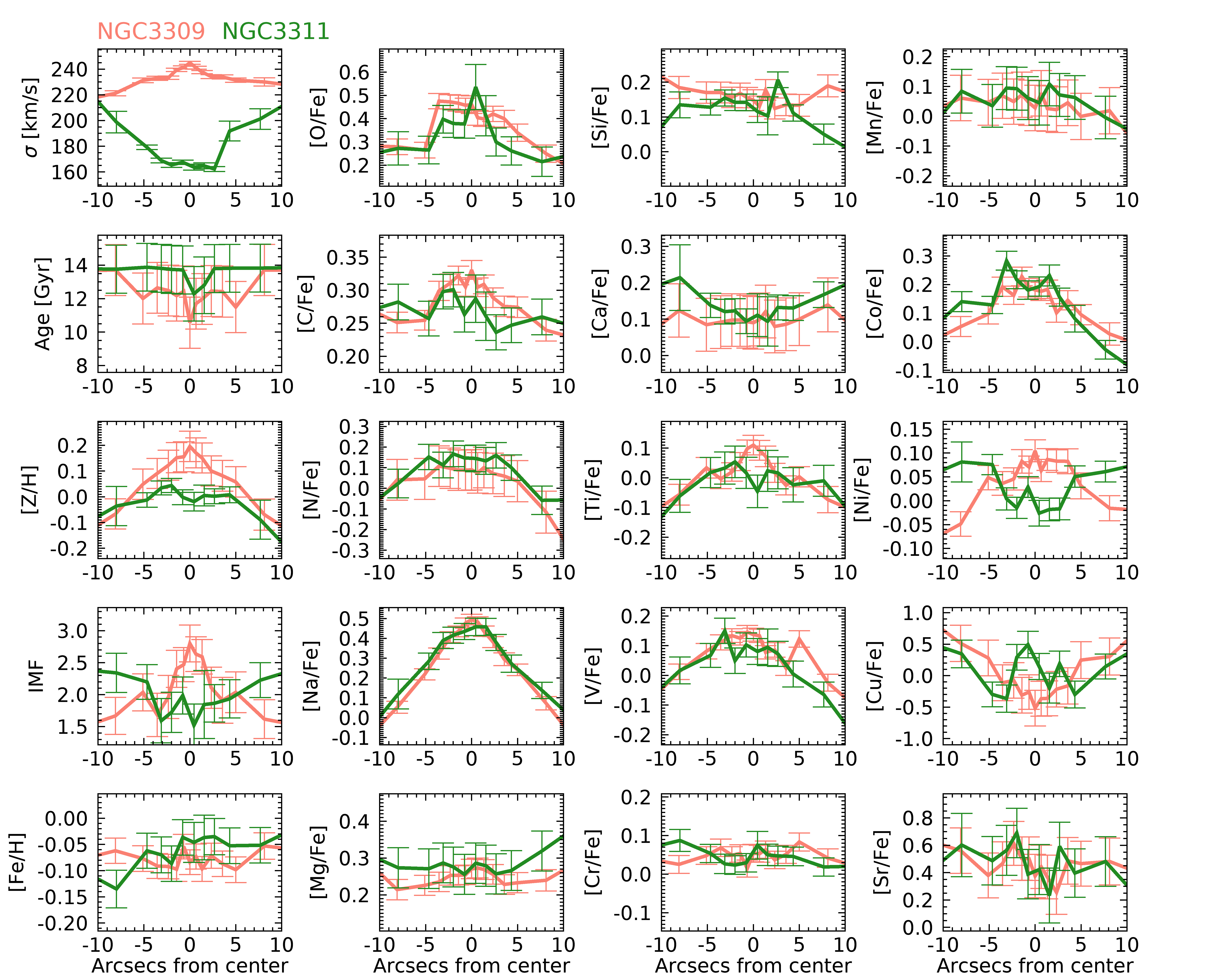}
\caption{\small{Zoom on the comparison of the retrieved parameters in the centers of the two galaxies: NGC3309 (pink) and NGC3311 (green). Excluding the velocity dispersion, metallicity and IMF, all the other parameters show very good agreement suggesting that the cores of the two companion galaxies have been formed from the same enriched material and at the same time. Differences in the velocity dispersion and IMF can be sign that  the IMF trend is connected with the kinematic distribution during galaxy formation.}}
\label{fig:comp2gal}
\end{centering}
\end{figure*}

The centers of elliptical galaxies are thought to host the core of the \textit{in situ} stellar population in the two phase formation scenario \citep{naab09,oser10}; thus focusing on the central regions of NGC3311 and NGC3309 may give us  hints of their formation history. To better visualize this comparison, in Figure \ref{fig:comp2gal} the central $20\arcsec$ of both galaxies for most of the retrieved stellar population properties are shown. This region corresponds to the limit of the deep potential well, just before the sharp rise of the velocity dispersion (see Figure \ref{fig:kinematics}). As can be seen in the upper left panel, and as already mentioned, the two galaxies show a significant difference in their velocity dispersion profiles, with NGC3309 (pink) showing the typical negative gradient of massive elliptical galaxies, and NGC3311 (green) instead exhibiting a flat trend in the center following a decrease from the outskirts. Other differences can be observed in the very center ($<2\arcsec$) where the metallicity and IMF (and Ni) both have  higher values in NGC3309. On the other hand, the age and all elemental abundances in these regions, within their uncertainties, have the same values. This is an important result suggesting that the stars in the cores of these two objects have formed at the same cosmic time and from a similar chemically enriched material.  

A number of studies have found correlations between the abundance patterns and velocity dispersion (e.g.: \citealt{worthey14,conroy14,parikh19,feldmeier21}), with generally more elemental enhancement at higher masses (see also Section \ref{sec:correlations}). 
Our two galaxies, however, deviate from these correlations. Indeed, their centers show similar abundances while their velocity dispersion is different. Moreover, also their dynamical mass is different, with M$_{dyn}^{NGC3309}/$M$_{dyn}^{NGC3311}$ $=0.7$, when derived both in the core and at R$_e$. This could be a particular case, though, since the proximity of the two stellar systems may suggest that both \textit{in situ} populations have effectively originated during the same star forming event. However, the difference in their IMFs suggests different paths in their star formation histories.

\subsection{Outer regions results}
While in the center of the two galaxies the parameters are precisely retrieved with well-shaped trends, in the outer regions and halos we find larger uncertainties and some scatter. As anticipated in Section \ref{sec:outer}, this can be connected to many factors: i) the lower S/N of these regions (Figure \ref{fig:SN}) due to the lower surface brightness $\mu>22$ mag/arcsec$^2$ in the outer $20-30\arcsec$ from the centers \citep{arnaboldi12}, ii) the increasing velocity dispersion that in $\sim20\arcsec$ doubles its value and broadens the absorption features causing more degeneracy among parameters, iii) the use of simple stellar population models on likely mixed stellar populations due to later accreted stars and minor mergers. Indeed, some level of scatter was noticed and commented in \citet{barbosa16} in the regions at R$>$1R$_e$ from NGC3311, and accredited to the presence of multiple components that make the halo around NGC3311 not homogeneous. 

Allowing for the above-mentioned caveats, we give a first estimate of the detailed chemical content of the stellar halo surrounding  NGC3311 and NGC3309 within the stated uncertainties. As shown in Figure \ref{fig:elements}, most of the elemental abundances in the halos show evident differences with respect to the two galaxy centers, with some going toward sub-solar (e.g. Fe, O, C, Na, Ca, Sr) or super-solar values (e.g. Mg, Si, Ti, V, Co, Ni). However, considering the large uncertainties of the outer regions, significant differences can be confirmed only for Fe, Na, Si, V, Co, Ni, Cu and Sr; even so, the last 5 elements from this list still suffer the lack of calibration corrective factors in models, which, as mentioned in Section \ref{sec:alfsetup}, are expected to be larger in regions of low metallicity. All $\alpha-$elements (C, O, Mg, Ca, Si, Ti) have rather similar trends, although, again, they show large error bars. 

Both higher S/N data and an improvement of models and codes to include mixed stellar populations would be required to significantly improve on this work. At present, we can provide a comparison with results of similar works to strengthen our findings.
Studies of elemental abundances in galaxy centers are rare in the present literature; however,  results in the outer regions are even rarer. We can compare our values in the outer regions with those from stacked spectra of \citet{greene15}, for Fe, Mg, C, N and Ca out to $\sim60\arcsec$. We find good agreement for Fe, Mg and C, and some deviation for N and Ca. Calcium in particular has more sub-solar values in our results, but the disagreement may simply result from the different way in which it is determined, i.e. from the optical Ca$4227$ index alone for \citet{greene15}. We note that fitting many Ca-sensitive features (like CaH, CaK, CaT) in our case, doesn't necessarily imply a better constraint (see Appendix \ref{app:gencorr}). However, our Ca is likely  well-constrained since we fit the many elemental abundances on which Ca features also depend. 
\citet{parikh19} studied the stacked spectra of early-type galaxies out to $1$R$_e$, extracting C, Mg, N, Ca, Na and Ti. Comparing with our results at $\sim1$R$_e$, we find mild consistency with C and N, and more evident deviations for Na, Ti, Mg and Ca. Also of note is that  the metallicity is different, with our values $\sim0.5$ dex more sub-solar, which may be due to the different adopted models. Other works retrieved fewer parameters (i.e. age, metallicity and some also [$\alpha$/Fe]), but extended to larger radii, as \citet{boardman17}, \citet{goddard17}, \citet{greene19} and \citet{perezhernandez22}. However, given the larger scatter in the parameters in the outer regions for all of these measurements, we conclude that further data and a more complex treatment of way to include mixed stellar populations in the comparison with models are warranted to better constrain the halo properties. Nevertheless, we note that stellar parameters derived from the western, central and eastern halo regions are generally centered on very similar values, suggesting that, despite their larger uncertainties, they hold a chemical identity and likely share their past accretion history.  We further discuss their possible origin in Section \ref{sec:halo_origin}.

\subsection{Correlations Among Measured Parameters}
\label{sec:correlations}

Correlations among physical parameters have been observed and studied in galaxies in order to find  the drivers and mechanisms of their star formation history \citep{maiolino19}. These correlations can quantitatively contribute to the improvement of galaxy formation and evolution models, as described, for example, in \citet{pipino09}, \citet{vincenzo16} and \citet{guidi18}.
Moreover, correlations among different chemical species can give us clues on the nucleosynthesis of each particular element (e.g.: \citealt{worthey14,maiolino19}). 

An often discussed global correlation is that between the stellar metallicity and the velocity dispersion $\sigma$ (e.g.: \citealt{trager2000,thomas05,gallazzi05,thomas10,mcdermid15}). Recently, it has been investigated if this correlation still holds within the same galaxy when measuring [Z/H] and $\sigma$ as a function of the galaxy radius, and also expanded to individual elements (e.g.: \citealt{worthey14,greene15,parikh19,feldmeier21}). Finally, with increasing indications that the IMF is not universal, a radial correlation of the low-mass IMF with $\sigma$ is also under investigation (e.g.: \citealt{conroy12,cappellari12,spiniello14}) but still debated \citep{barbosa21b,feldmeier21}. Since our galaxies have two different velocity dispersion profiles, particularly in their inner regions, our results are an optimal benchmark to test the global validity of such correlations.

However, before investigating possible correlations among stellar properties, it is necessary to take into account the correlations that occur during the fit among parameters, i.e. their degeneracy. To do this, we inspected the marginal posterior distributions of all the pairs of parameters for each analyzed spectrum, calculated the Spearman coefficient and took into account those relevant in the discussion below in  Sections \ref{sec:imfcor} and \ref{sec:scalingrelation}. All the details of the fit correlations analysis are described in Appendix \ref{app:fitcorr}.

In the following sections we will focus on specific correlations, i.e. those with the IMF slope and velocity dispersion, and leave the comments on more general correlations to Appendix \ref{app:gencorr}.

\subsubsection{Correlations with the IMF}
\label{sec:imfcor}

Among radial correlations, those with the IMF slope are actively under study, for example, in \citet{sarzi18}, \citet{barbosa21b} and \citet{gu22}. In particular, \citet{barbosa21b} have shown for our same galaxy NGC3311, that a robust radial correlation with the IMF can be found with the age and not with $\sigma$. Although observed, a correlation with [Z/H] was not considered reliable by these authors since they observed a similar positive trend in the posterior-distribution, thus addressing the correlation to internal fit degeneracy (as we discuss in Appendix \ref{app:fitcorr}).
We notice that along our long-slit direction, our {\it IMF slope} is fully consistent with the values of \citet{barbosa21b} showing a flat trend, although 
their overall distribution of values as a function of radius from all the Voronoi bins around NGC3311's center presents a mild negative gradient.
On the contrary, and similarly to their previous work \citet{barbosa16}, we do not see the same sharp negative {\it age} gradient, which in \citet{barbosa21b} is rather significant.  
Our results are consistent with the positive IMF-[Z/H] correlation and we checked that our internal fit degeneracy has only a mean $\rho=0.21$ with $p=0.13$, indicating a low probability of finding a correlation due to degeneracy.

These differences underscore the difficulty of comparing results of stellar populations from analysis based on different codes and models. With our analysis, albeit based on only one direction across the galaxies, we can compare the two companions in a robust way. Moreover, in Appendix \ref{app:imf} we discuss the strengths of our IMF measurements, including the comparison with the expectations of spectral indices.

\begin{figure*}[ht]
\begin{centering}
\includegraphics[width=8.7cm]{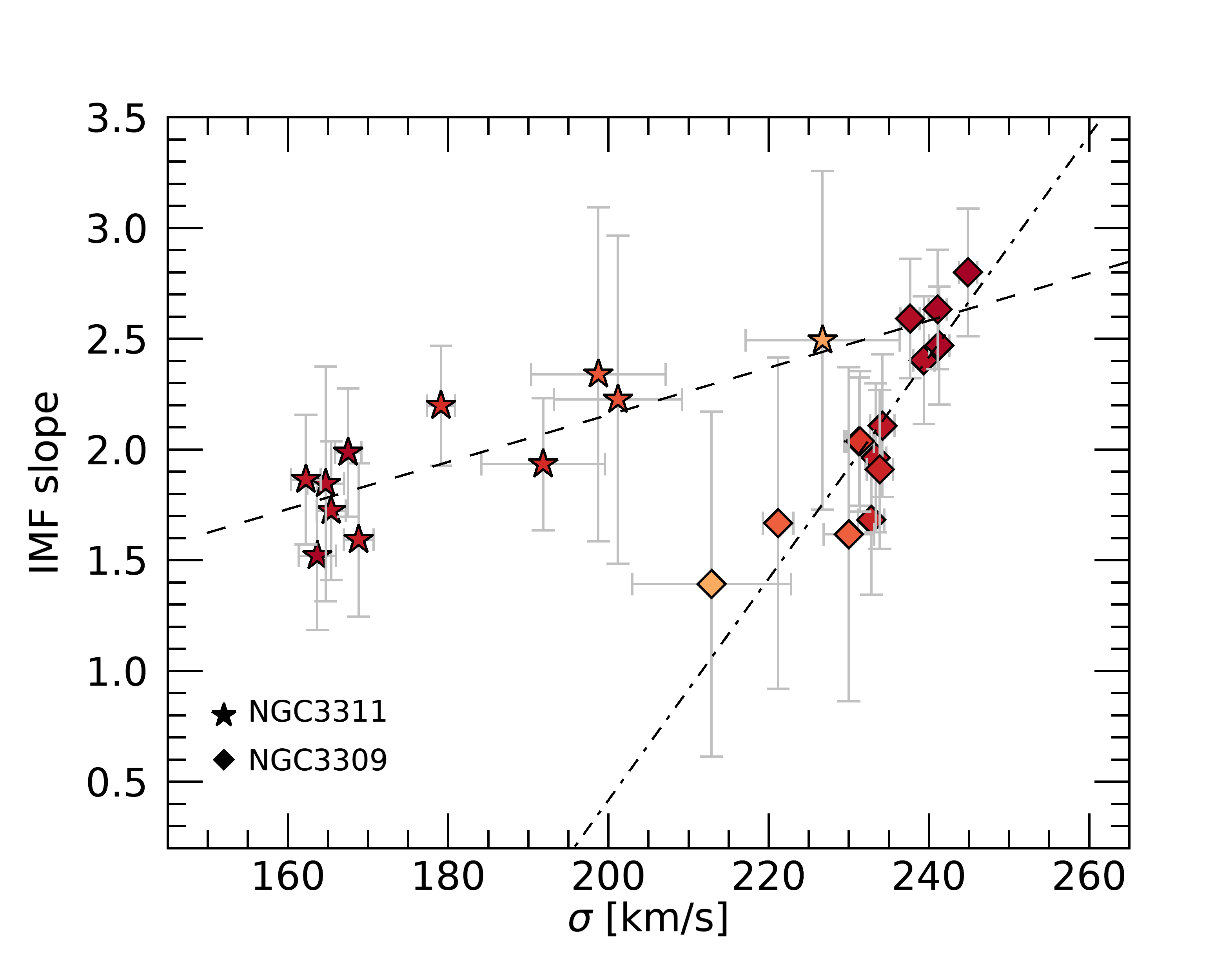}
\includegraphics[width=8.7cm]{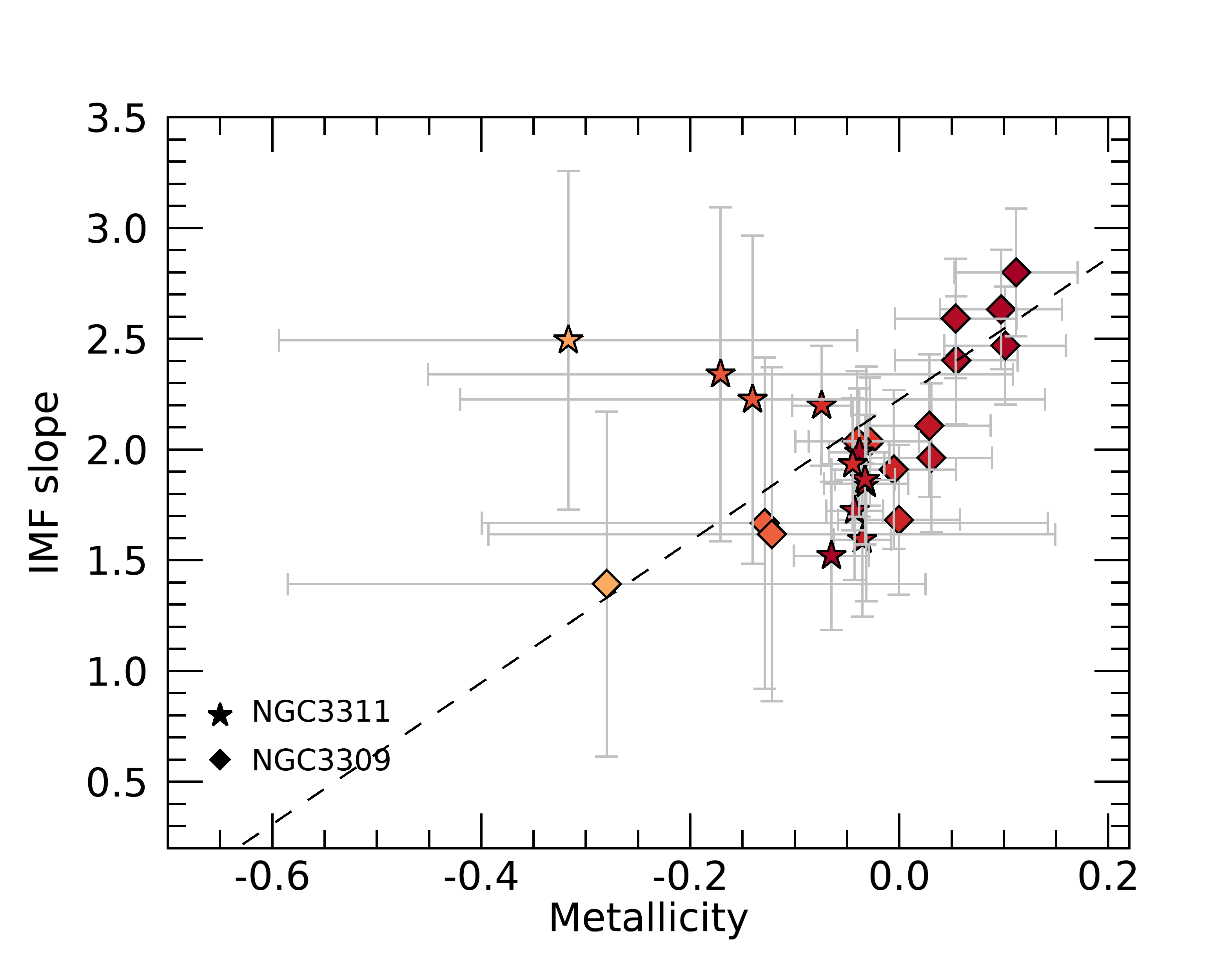}
\caption{\small{Correlations between IMF slope and velocity dispersion (left), and metallicity (right). Points are color-coded as in Figure \ref{fig:ageZimf}, with darker red in the center. Diamonds refer to NGC3309 and stars to NGC3311. On the left, the dashed line indicates the linear fit to the NGC3311 points and the dot-dashed line the fit to those of NGC3309; on the right, the dashed line refers the global linear fit.}}
\label{fig:correlations}
\end{centering}
\end{figure*}

We show the trends of the IMF slope with $\sigma$ and metallicity in the centers of these galaxies in Figure \ref{fig:correlations}. 

Regarding the dependence of the IMF on the velocity dispersion (left panel), we first compare our results with the global relations from, for example, \citet{conroy12} and \citet{cappellari13}. By averaging our central values, we indeed find a good consistency with their findings.
However, our two local IMF slope trends follow two different positive correlations as shown by the line in the left panel of Figure \ref{fig:correlations}: a much steeper one for NGC3309 (diamonds, dot-dashed line), and a milder one for NGC3311 (stars, dashed line) that is also consistent with a flat trend. Interestingly, higher values of both $\sigma$ and IMF slope are seen in the center of NGC3309 and, on the contrary, in the outskirts of NGC3311, as a consequence of their opposite radial profile of both $\sigma$ and IMF.

As  noted in the following Section \ref{sec:scalingrelation}, generally the global scaling relations are not equally replicated by the local ones. In addition, \citet{parikh18} show the local trend of the $\alpha$ mass excess factor with $\sigma$ and find different slopes with respect to the global relation. With the direct comparison our two objects , we further show the complexity and peculiarity of each galaxy, but also confirm that a trend with $\sigma$ holds for both of them. 
We thus conclude that, although there is not a unique trend that radially correlates  the IMF slope with the velocity dispersion in an absolute way, $\sigma$ and the IMF may be interconnected and local processes may also affect this relation.

IMF trends with metallicity of both galaxies show a positive correlation (Figure \ref{fig:correlations}, right panel, dashed line). Few outliers may be explained by the higher uncertainties in the IMF measures in the outer regions, as described in Appendix \ref{app:imf}. 
While the correlation for only NGC3309 is strong with $\rho=0.88$ and $p=0.000025$, the overall correlation is milder with $\rho=0.41$ and $p=0.04$. However, the most central values of NGC3311, where errors are smaller, are well fitted in the trend, reinforcing the hypothesis that a local connection between IMF and metallicity does hold.
This finding confirms the results of \citet{parikh18} who find that regardless of the mass bin or radial position, the IMF tracks very well the total metallicity globally and locally in a similar way. Other examples of similar results come from \citet{martin15c}, \citet{vandokkum17} and \citet{feldmeier21}.

We have also checked for the presence of a correlation between IMF and [Mg/Fe] in the two centers, but found no correlations, neither separately, nor together. This is in agreement with \citet{labarbera15} and \citet{martin15c}. Indeed, we stress once again, that the retrieved single-element abundances are very similar in both galaxy centers.

In general, as noticed in \citet{vandokkum17}, the correlations of the IMF slope with single elemental abundances show larger scatter and differences; only the overall metallicity follows the IMF trend. 

\subsubsection{General scaling relations} 
\label{sec:scalingrelation}

In Figure \ref{fig:correlationsALL1} and \ref{fig:correlationsALL2}, all elements as a function of the velocity dispersion are shown, with the same color-coding of Figure \ref{fig:ageZimf}, i.e. with darker red in the two galaxy centers and bluer in the outer halo regions. In the lowest rightmost panel of Figure \ref{fig:correlationsALL2}, the [$\alpha$/Fe] trend, calculated as the average of C, O, Mg, Ca, Si and Ti, is shown. 

A general comment is that the well-established global scaling relations among early-type galaxies are not easily reflected in the local relations, as also observed in \citet{parikh19}, here complicated also by the presence of the outer stellar halo. This is evident for the [Z/H] vs $\sigma$ relation, that in our results shows a general clear negative gradient, while we usually find higher metallicity at higher velocity dispersion. We remark that by fitting the spectrum composed by stacking all the radial regions spectra within 1Re of each galaxy, thus considering the observation of the global galaxy, we obtained results perfectly in agreement with local relations \citep{thomas10}, with a (slightly) higher metallicity for a (slightly) higher velocity dispersion ([Z/H]$_{NGC3311}=-0.13\pm0.04$ dex and $\sigma_{NGC3311}=215\pm5$ km/s, and [Z/H]$_{NGC3309}=-0.04\pm0.01$ dex and $\sigma_{NGC3309}=238\pm2$ km/s).
Our result indicates that locally within galaxies, the [Z/H]-$\sigma$ positive correlation is only an artifact that occurs because typical elliptical galaxies have a decreasing velocity dispersion profile. As in the case of NGC3309's center, indeed, we find a steep increasing local gradient of [Z/H] with $\sigma$ as a consequence of their both negative gradient with R. This is also valid for some of the elemental abundances as Na and Ti (see Figure \ref{fig:correlationsALL1}). 

While in the centers, elements behave smoothly and homogeneously, with more flat trends for the lower and flatter velocity distribution galaxy (NGC3311) and steeper positive gradients with $\sigma$ for the other one which has a steeper velocity distribution profile (NGC33309), if we look globally, trends with $\sigma$ are more difficult to be spotted and justified. At $\sigma>250$ km/s, the distinction between the two galaxies is erased, meaning that the surrounding halo is equally non-homogeneous. 

In these areas elements exhibit more scatter but, with the exception of Ni and Ti, values are or all sub-solar or all super-solar. This could be the first sign that the halo regions have their own unique chemical identity, although the large scatter decreases the statistical significance of this result.

Moreover, it must also be noted that in these outer regions and halos, the higher velocity dispersion values are not tracing higher stellar mass contributions as in the central regions, but a higher contribution from dark matter. Indeed, \citet{richtler11}, analyzing the velocity dispersion profile derived from NGC3311 and its surrounding globular clusters out to $\sim200$ kpc, have argued that a cored dark matter halo is necessary to explain the observed kinematics.

\begin{figure*}[ht]  
\begin{centering}
\includegraphics[width=16.cm]{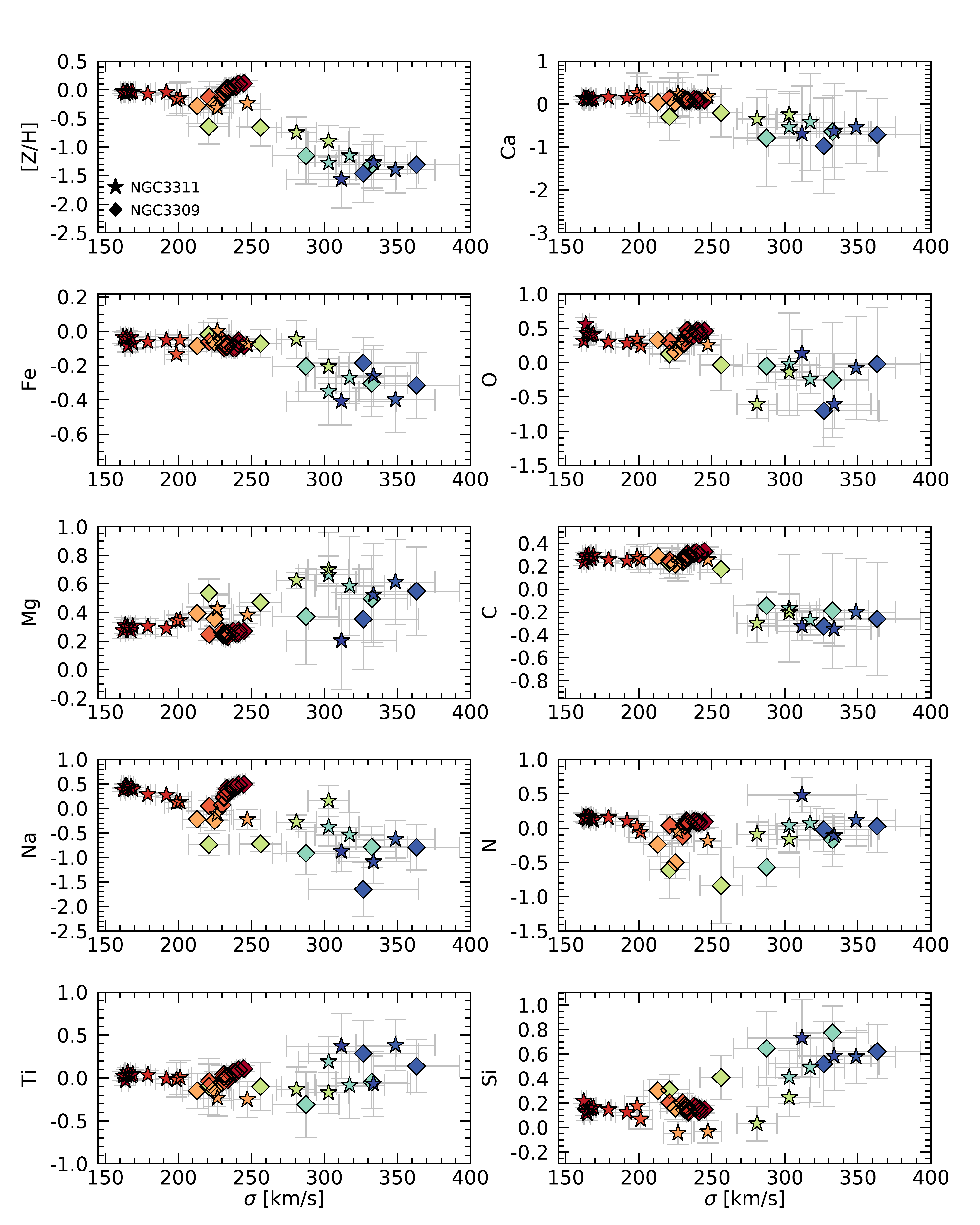}
\caption{\small{Correlations with velocity dispersion of metallicity and all retrieved elements. Points are color-coded as in Figure \ref{fig:ageZimf}, with darker red in the center and blue in the outskirts. Diamonds refer to NGC3309 and stars to NGC3311. Within $250$ km/s, in the central regions, it can be easily seen that the different $\sigma$ radial profiles of the two galaxies produce different trend of elements with $\sigma$, proving that local scaling relations are only explained by their $\sigma$ profiles. At $\sigma>250$ km/s, instead, elements in the halos are distributed in a different way, which is hard to connect to the inner regions due to the  increasing contribution of dark matter.}}
\label{fig:correlationsALL1}
\end{centering}
\end{figure*}
\begin{figure*}[ht]  
\begin{centering}
\includegraphics[width=16.cm]{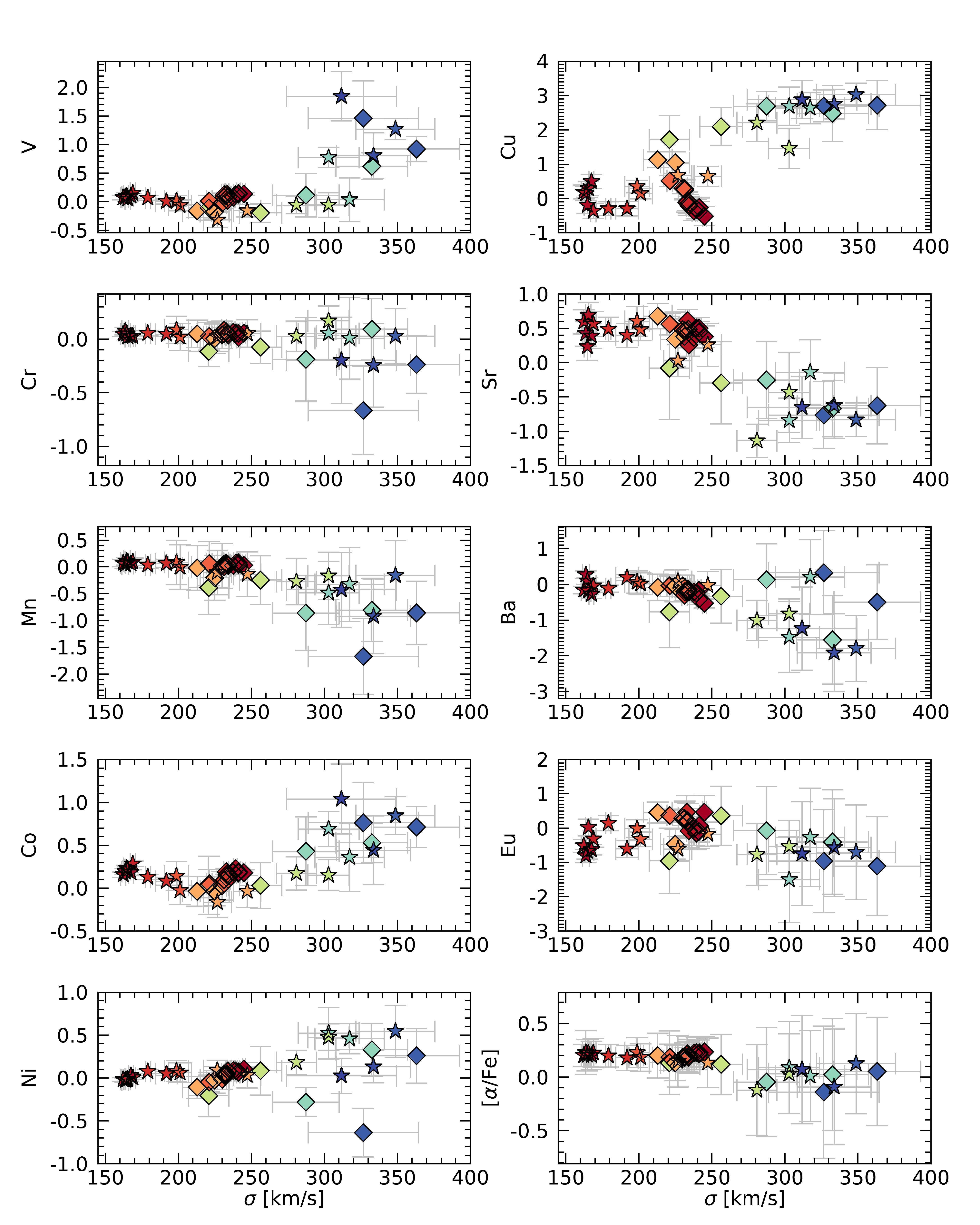}
\caption{\small{Continued.}}
\label{fig:correlationsALL2}
\end{centering}
\end{figure*}

\subsubsection{$\alpha$-elements}
\label{sec:alphafe}

The global scaling relation of [$\alpha$/Fe] with $\sigma$ has been also widely observed and discussed (e.g.: \citealt{thomas05,thomas10,johansson12}), thus far indicating an enhancement of $\alpha$-elements in more massive galaxies. This parameter has been studied in particular because it is directly connected with the timescale of the star formation process \citep{thomas05}. Specifically, more massive galaxies, observed to be more $\alpha$-enhanced, are thought to have formed with a shorter timescale, typically $<1$ Gyr for early-type galaxies. 

As a consequence, considering the centers of NGC3309 and NGC3311, and that the ratio of their dynamical mass is $\sim0.7$, we would expect a difference in the $\alpha$-elements abundance. Instead, if we average all the retrieved $\alpha$-elements together, we find consistent values between the two galaxies. Indeed, in Section \ref{sec:centers} and Figure \ref{fig:comp2gal} we have already shown the high similarity of elemental abundances in the two central regions.
The same flat [$\alpha$-Fe] vs $\sigma$ trend, as shown in Figure \ref{fig:correlationsALL2}, remains constant also out to higher $\sigma$ values, i.e. in the halos (as also observed in \citealt{barbosa16}). 

As already noted in \citet{graves08}, and widely investigated afterwards, the abundance patterns of stellar populations are too complex to be described with only one parameter. Indeed, the detailed abundance characteristics of stellar populations offer a wealth of information on galaxy formation processes and stellar nucleosynthesis. With the possibility of observing the radial variation of many single $\alpha$-elements, we can thus hope to better characterize the past history of these cluster members. 

\begin{figure*}[ht]  
\begin{centering}
\includegraphics[width=18.cm]{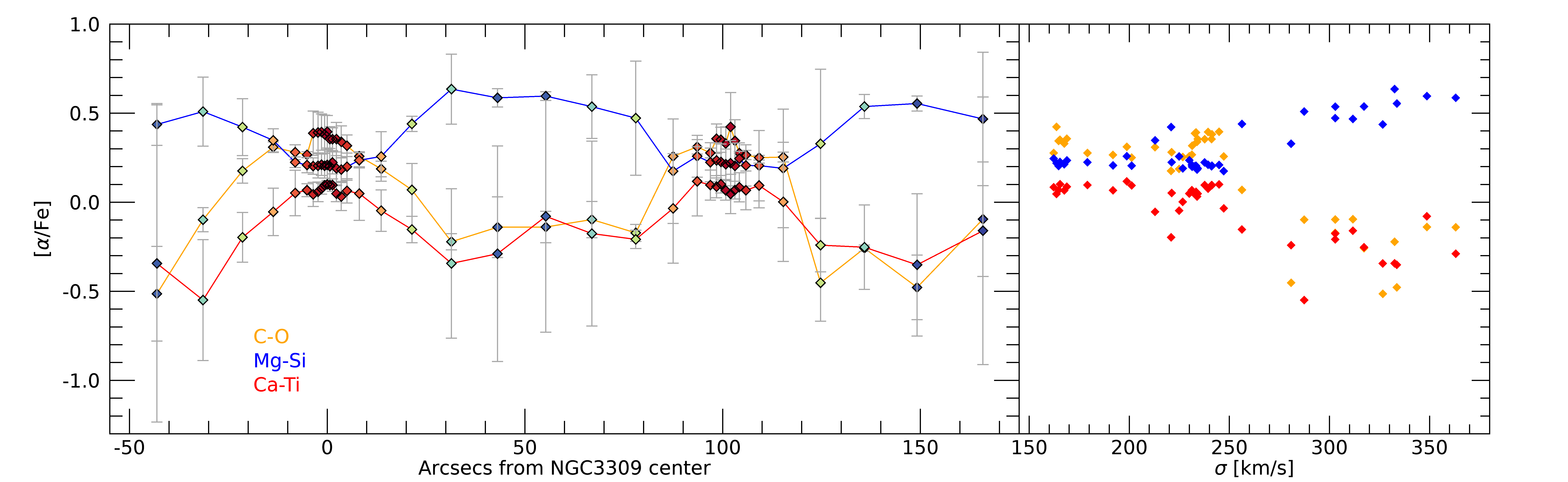}
\caption{\small{[$\alpha$/Fe] trend when averaging: C-O (orange), Mg-Si (blue) and Ca-Ti (red). Left panel as a function of the distance from NGC3309, right panel as a function of the velocity dispersion. Error bars in the left panel are the standard deviations of each average; to better see the trends, error bars are omitted in the right panel.}}
\label{fig:alphafe123}
\end{centering}
\end{figure*}

We then derived the [$\alpha$/Fe] trend in three different ways by averaging C-O, Mg-Si and Ca-Ti separately. These trends are shown in Figure \ref{fig:alphafe123}, from which it is clear that they do follow different behaviours, as expected. C-O (orange) traces roughly the Ca-Ti (red), albeit being higher mostly in the centers, while Mg-Si (blue) shows a totally different behavior especially in the halos. In the right panel of Figure \ref{fig:alphafe123}, with the same colors, we show the three [$\alpha$/Fe] trends as a function of the velocity dispersion. It can be noticed that the Mg-Si trend clearly shows a positive gradient, totally in contrast with the other two trends. 

If we would have considered the average of all $\alpha$-elements trend alone, as is usually done, we would have concluded that the central regions and halos shared the same $\alpha$-enhancement and thus star formation timescale.
Instead, by inspecting separate elements, produced by different processes or by a different mix of them, it becomes clear that the halo regions have probably experienced a different star formation history. If following the [Mg/Fe] ratio, as used in \citet{thomas05,thomas10}, our Mg-Si trend suggests a star formation timescale in the range $0.2-1$ Gyr for regions within $1$R$_e$, and $<0.1$ Gyr in the outer regions. 

We have also checked if the difference among $\alpha$-element trends can be addressed to the corrections applied during the post-processing of the fit. Without the corrections the difference is still visible, however the Mg-Si trend is flat at $\sim0.2$ dex also in the halo regions, with the consequence of an overall constant star-formation at $\sim1$ Gyr.  

From the analysis of the detailed $\alpha$-elements we can conclude that: i) regardless of their different mass and velocity dispersion, the core of the two galaxies have formed with the same star formation timescale, ii) the outer regions show signs of different production mechanisms for different $\alpha$-elements, iii) the outer regions stars have formed with a different star formation timescale than the centers.

\subsection{The possible origin of halo stars}
\label{sec:halo_origin}

In Section \ref{sec:outer} we have discussed the similarity among the western, central and eastern regions. With the results obtained on the $\alpha$-element distributions and the related information on the star formation time-scale discussed in Section \ref{sec:alphafe}, we can also speculate on the possible origin of the halo stars. 

The relatively small spread of elemental values around the cluster seems consistent with an origin due to the accretion of dwarf galaxies rather than globular clusters,  the latter of which, with their significantly larger numbers, would have resulted in a more diverse distribution of abundance values \citep[e.g.,][]{aoki20, ji2020}.
Moreover, the single $\alpha$-element trends in the halos (see Figure \ref{fig:alphafe123}) show a clear difference in their values in the central regions, suggesting that the stars in the halos have been formed in different stellar systems, where the star formation occurred over different time-scales. A possible origin of such stars could be dwarf spheroidal or ultra-faint dwarf galaxies. Indeed, these kinds of stellar systems are known to host chemical properties similar to the ones we observed in the Hydra halo, i.e. very low metallicity associated with, for example, super-solar [Mg/Fe] and very low [Ba/Fe] \citep{koch08b}.

Additional observational evidence supporting this scenario can be found in the works of \citet{coccato11}, \citet{ventimiglia11} and \citet{arnaboldi12}. On the basis of stellar population parameters and kinematic properties of planetary nebulae in the halo around NGC3311 and NGC3309, these works show evidence of accreted satellite galaxies which have been tidally stripped and then diffused into the stellar halo. 
This mechanism is witnessed to be still ongoing for the dwarf low-metallicity galaxy HCC 026 and the S0 galaxy HCC 007, in the proximity of the cluster center, where tidal streams are observed and dynamically characterized.


\section{Summary and Conclusions}
\label{sec:conclusions}


In this work we have analyzed high-quality long-slit optical+NIR spectroscopic data across the two brightest galaxies of the Hydra I cluster. 
We have characterized in detail the stellar population of their centers, where the \textit{in situ} component still resides, and compared them to their surrounding stellar halos where, on the contrary, their evolution has resulted in mixed stellar components.

The advantage of studying these two galaxies together is that we have been able to compare directly data taken with the same instrumental setup and also using the same data-reduction methods and stellar-population-synthesis models. In addition, we have been able to test the validity of many scaling relations (in a local setting), since the two objects differ in their mass and velocity dispersion radial profiles.

With full spectral fitting over a large wavelength range, in comparison with stellar population synthesis models allowing for non-solar values of elemental abundances and a non-constant IMF slope, we determined age, overall metallicity, IMF slope and $19$ elemental abundances with good precision. Due to the lower signal-to-noise of IMF-sensitive spectral features in the halo regions, the IMF could only be robustly derived  in the centers of the two galaxies. 

Despite their different masses and velocity dispersions, we find that the two galaxy centers are very similar in their stellar content, with same age and same elemental abundances. This suggests that their formation happened at the same cosmic epoch and that they shared a similar chemical enriching history. Moreover, since we can correlate $\alpha$-elements with the star formation time-scale, it also appears that their star formation history has been prolonged in the same way.

Beyond such shared characteristics that may suggest that NGC3311 and NGC3309 followed a similar evolutionary path, we also measured some disparities that suggest a slightly more complex picture.
In particular, the two galaxies differ in their overall metallicity, the IMF slope, and the radial velocity dispersion profile.

Focusing on these three properties that change, and investigating their possible relation, we found that: i) the IMF correlates well with [Z/H] both locally and globally, with higher metallicity having a bottom-heavier IMF.  
Although the difference in metallicity in the centers of the two galaxies is small ($\sim0.1$ dex), the \textit{local} metallicity-IMF correlations are consistent with the suggestion by  \citet{martin15c} that the metal content could have affected the initial collapse of the molecular clouds,
thus shaping  the low-mass end of the IMF. We can also speculate that the high-mass end of the IMF, co-responsible for the chemical enrichment, is similar for the two galaxy centers.
ii) The IMF correlates with the velocity dispersion, with higher $\sigma$ connected with a bottom-heavier IMF. Moreover, we found that the local correlations of the two galaxies have different slopes, suggesting not only that the IMF and $\sigma$ can be related, but also that local processes within the same galaxy can drive this connection.

In a similar way, the elemental abundance trends with $\sigma$ also show different local behaviours for the two galaxies. Given the different velocity dispersion profiles of NGC3311 and NGC3309,  we were able to distinguish trends that are likely robust global correlations from those that are only the consequence of possessing both abundance gradients and a negative velocity dispersion gradient, typical of elliptical galaxies.

Analysing the outskirts and stellar halo regions, we found gradually larger uncertainties in the retrieved stellar properties. These larger uncertainties are due to many factors, including a lower signal-to-noise, as well as the increasing line broadening due to the higher velocity dispersion and the presence of mixed stellar components. Indeed, the investigation of stellar halos is at the moment limited by the lack of fitting codes that allow for multiple populations that differ not only in their age but also in their chemical properties. 

In addition to these limits, we found clear chemical patterns in the halos, with homogeneity among the eastern, western and the regions between the two galaxies, suggesting an overall common evolution for the central $\sim200$\arcsec\space of the cluster. Although it is not yet possible to resolve stars in such distant stellar halos, a dedicated study of chemical properties for different regions around the Hydra cluster halo would help in understanding the origin and nature of the accreted systems. From our findings we can speculate that the origin of the halo stars can likely resides in dwarf spheroidal or ultra-faint dwarf galaxies that have been bounded to the cluster potential well, as also observed in previous works \citep{ventimiglia11, arnaboldi12}.

Further investigations will be needed to confirm these findings and in particular to understand if they are a characteristic of cluster galaxies.

\acknowledgments
We are thankful to the anonymous referee for reviewing the manuscript and for the helpful suggestions. I.L. thanks B. Madore for his help during the observations of the data used in this work.
This research has made use of ``Aladin sky atlas'' developed at CDS, Strasbourg Observatory, France.
The Wisconsin H-Alpha Mapper and its Sky Survey have been funded primarily through awards from the U.S. National Science Foundation.

%

\vspace{5mm}
\facility{Magellan: Baade (IMACS)}
\software{IRAF (Tody 1986, Tody 1993), IDL, MOLECFIT (Smette et al. 2015; Kausch et al. 2015), emcee (Foreman-Mackey et al. 2013), ALF (Conroy et al. 2021, 2018), PPXF (Cappellari 2017).}





\clearpage
\appendix
\section{Foreground emission}
\label{app:foreground}

\begin{figure}[ht!]
\begin{centering}
\includegraphics[width=15.0cm]{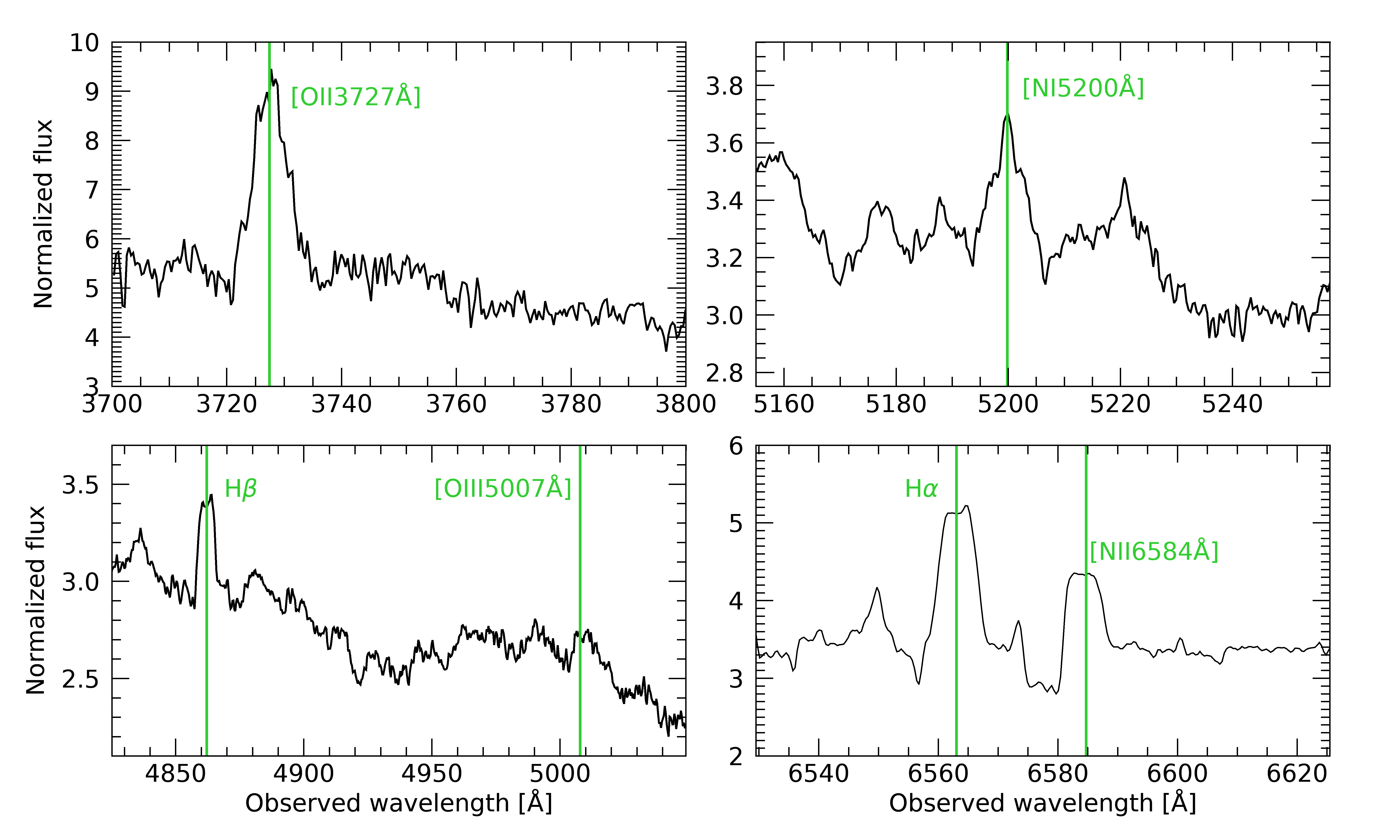}
\caption{\small{Foreground emission lines in four different regions of an example spectrum extracted in the halo between NGC3309 and NGC3311.}}
\label{fig:foregroundemission}
\end{centering}
\end{figure}

All the spectra extracted from the long slit positioned across the Hydra I cluster center show a uniform emission of the lines [OII$3727$\AA], H$\beta$, [OIII$5007$\AA], 
[NI$5200$\AA], H$\alpha$ and [NII$6585$\AA]. By analysing this emission, in particular the strongest one, i.e. [OII], it is clear that they do not belong to the cluster's 
light, but rather are local in the Milky Way.  We fit all the lines with a Gaussian profile, and found a mean cz$\sim25$ km/s with a mean velocity dispersion of $360$ km/s.
This foreground emission is constant along the whole physical direction of the slit.
Fig. \ref{fig:foregroundemission} shows a zoom of each foreground emission line in an example spectrum extracted in the halo between NGC3309 and NGC3311.

The observed lines and their intensities are consistent with those typical of the Warm Ionized Medium (WIM), as studied over the past two decades ({\it e.g.,} see the review by \citealt{haffner09}).
Indeed, the WIM is characterized by strong [NII] and [SII] lines, while [OIII] and [SIII] are generally weaker \citep{mathis00}. Moreover, [OII$3727$\AA] is generally the strongest line in the optical region. We also checked some line ratios, like [NII$6583$\AA]/H$\alpha$, and found that they are consistent with \citet{mathis00}. 
Generally, the WIM is considered to be produced by UV bright O-B stars, but studies are still investigating if other sources and mechanisms, like supernovae remnants \citep{raymond92}, or shock excitation \citep{martin97}, or dust scattered radiation \citep{barnes15}, can contribute to the ionization.

We found a confirmation of our detection of WIM in the foreground region of the Hydra I cluster from the Wisconsin H-Alpha Mapper Sky Survey (WHAM, \citealt{haffner03}). 
The WHAM survey scanned the whole sky at the H$\alpha$ wavelength with a spectral resolution of $12$ km/s and a spatial resolution of one degree. 
By inspecting the WHAM H$\alpha$ intensity map around the region of the Hydra I cluster, we confirmed that there is a moderate emission of H$\alpha$ which is constant around the cluster, as measured in our spectroscopic data.

With the aim of finding possible known candidates for the source of this emission, we looked for O-B-A stars around the cluster. We cross-checked the list of nearby 
UV-bright stars with Galex data \citep{bianchi11} using Aladin \citep{aladin2000}. We only found two stars with high flux in B-band, i.e. HD-91209 and HD-93657 with spectral type A3IV and A1V respectively,
but their locations do not match the morphology of the higher intensity regions in the {H$\alpha$} map very well. We have also checked for the presence of supernovae remnants from the 
catalog \citet{green19}, but did not find any matches. No visible X-ray emission sources were found in this same region: this is not unexpected since the X-ray emission originates from the potential well of the cluster at z$\sim0.013$, as spectroscopically confirmed by the Chandra data of \citet{hayakawa04}.

We further analyzed the photometric data available in both the B (Bessell-B1) and I (CTIO-I1) band, as observed with IMACS during the same night of the spectroscopic observations.
The large field of view of $15$\arcmin\space could potentially help in localizing  a hypothetical excess of blue light. We then reduced the photometric data in both 
bands and generated the color B-I frames of the Hydra I cluster. No blue excess is observed; the color is highly uniform in all observed regions.
 
We conclude that with the available data and catalogs, we are not able to retrieve the actual source of this diffuse ionization.


\section{Gas emission}
\label{app:gas}

\begin{figure*}[ht]
\begin{centering}
\includegraphics[width=10.5cm]{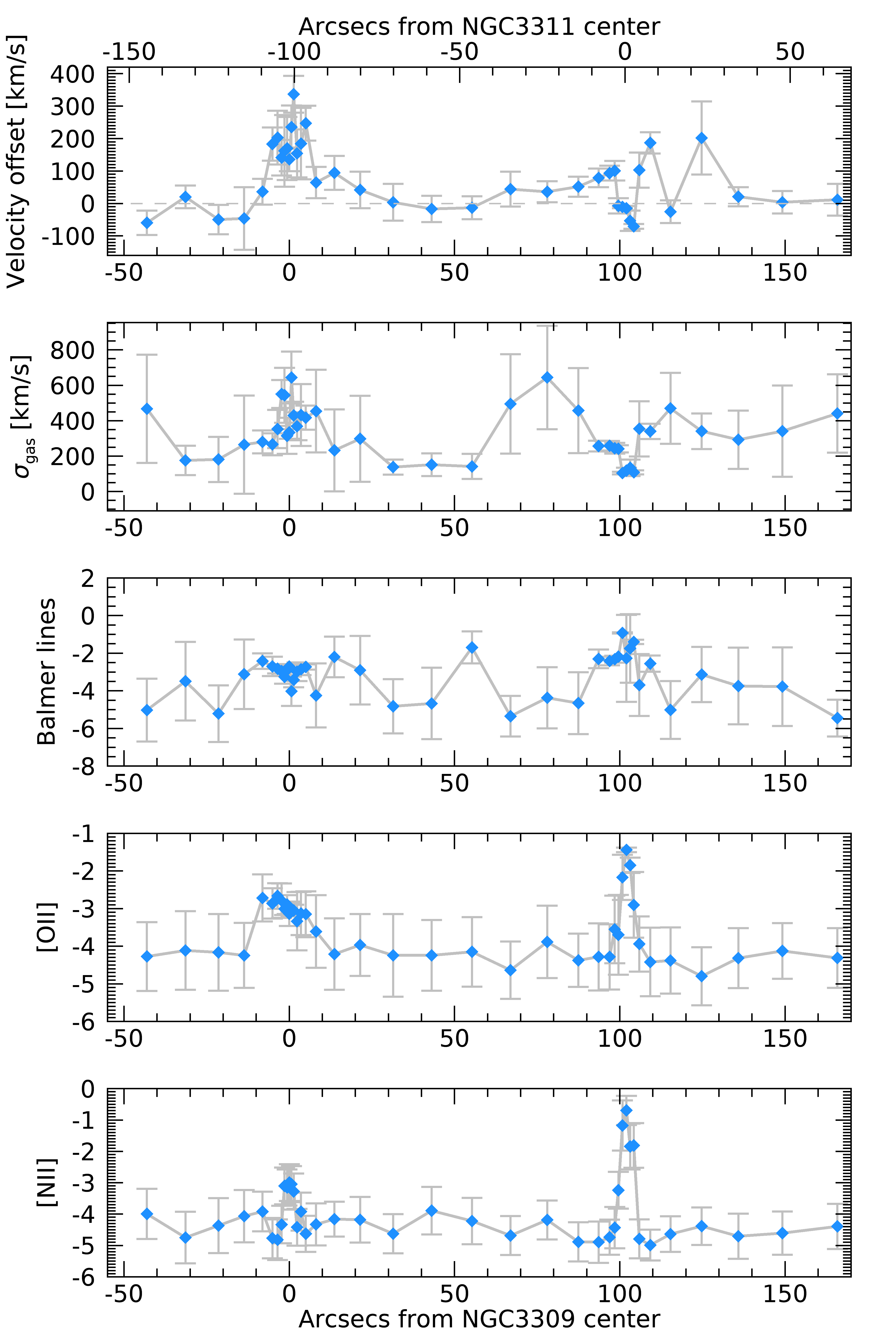}
\caption{\small{Retrieved gas emission properties. From the top to the bottom: recession velocity offset from the main stellar component, gas velocity dispersion, 
intensity of Balmer, [OII] and [NII] emission lines. 
}}
\label{fig:gas}
\end{centering}
\end{figure*}

In Figure \ref{fig:gas}, top two panels, the kinematics of the gas component, retrieved with ALF, is shown. The top panel indicates the relative shift in radial velocity 
of the gas component with respect to the main stellar component. We note that, for NGC3309, the small amount of gas present in its center is strongly decoupled from the 
stellar component with velocities offset by up to $300$ km/s. While for NGC3311, which hosts strong emission lines in its center (as shown for Balmer lines and [NII] 
in the lower panels of Figure \ref{fig:gas}), there is a velocity offset of only $100$ km/s from the stellar system. This is consistent with the findings of \citet{richtler20} (see their figure $6$), when considering the position of our slit with respect to the galaxy (see Figure \ref{fig:imaging}).

A difference between the two galaxies can be observed also 
looking at the velocity dispersion of their gas component: in the center of NGC3311 the gas has similar velocity dispersion values (comparing to Figure \ref{fig:kinematics}) to the 
stellar component, while NGC3309's center present much higher $\sigma$ values. The latter, however, have larger error bars probably due to the more difficult detection of the gas in emission considering its paucity. 

As specified in Section \ref{sec:alfsetup}, in the fit with ALF we included the emission of the following lines, typical of young, star forming regions, as 
free parameters: Balmer lines, [OII], [OIII], [SII], [NI] and [NII]. In Figure \ref{fig:gas} we only show the retrieved intensity of Balmer lines, [OII] and [NII], which are the strongest in our fitted wavelength range, and constrained best. Due 
to their weakness,  [OIII], [NI] and [SII] generally span all the possible values and have large error bars.

The detection of emission lines in the core of NGC3311 is not unexpected due to the presence of the dusty disk as shown in Figure \ref{fig:imaging}, as extensively observed in the literature (e.g. \citealt{lindblad77,wirth80,vasterberg91,grillmair94}). More recently, \citet{richtler20} showed that this inner dust disk embeds an ongoing star formation region whose gas, detected in emission, is perfectly confined within its limits. Taking into account the likely presence of a small and young stellar component in this confined region, we have fit two stellar components as allowed by ALF. The age of the minor one is bound in the fit from $0.5$ to $3$ Gyr. We found young component fractions $<1$\% with ages from $0.8$ to $1.3$ Gyr, in agreement with the estimates in \citet{richtler20}.
Regarding the center of NGC3309, we fit only the very central bin with a double component, since the retrieved age does not converge well otherwise. We found a $1.1$\% young component with an age of $1.8$ Gyr.

\section{Outer regions fit details}
\label{app:outer}

In this section we focus on how we treated the analysis of the outer regions and halos, located at $>10\arcsec$ and characterized by lower S/N, higher velocity dispersion, and likely mixed stellar populations.

Lower S/N and higher velocity dispersion make the absorption line features less evident and broader. Moreover, the large number of rejected pixels due to increasing sky residuals, mostly around $6400$\AA\space and $8400$\AA, makes the fit cover wavelength ranges over several gaps. Mixed stellar populations can instead create a bias on the retrieved parameters obtained with SSP models.

After several tests aimed at verifying the robustness of the fit on these outer spectra when changing wavelength region and using ALF in either full or simple mode (the latter consisting of a smaller set of free parameters), we noticed that the retrieved parameters that suffer most from the above-listed difficulties are the kinematics ones, i.e. the velocity dispersion and the radial velocity. Indeed, these parameters did not converge well even after increasing the number of walkers.    

Since the kinematics severely affect the retrieval of stellar population parameters  due to degeneracy, we further investigated this problem as follows. We fit each spectral range separately with ALF in the super simple mode (retrieving only the kinematics, age and metallicity) and compare the kinematics results with those obtained with PPXF, set up as described in Section \ref{sec:data}. The comparison shows large scatter both among results obtained with the same code but from different spectral ranges, and between the results obtained with different codes on the same wavelength region. The only observed good consistency between the two codes was the results from chip 5 (around $5500$\AA) and partially chip 8 (around $4700$\AA), where the spectra have higher signal and a larger number of strong features (e.g. Mg$_b$). Repeating the same test on inner galaxy spectra instead, we found a full consistency.  

Kinematic results from chip 5 are also in very good agreement with those in the literature, as detailed in Section \ref{sec:results} and shown in Figure \ref{fig:kinematics}. For these reasons, we decided to fix the kinematic values of the halo and outer regions to those obtained by our ALF run on chip 5 spectra with the super simple mode.

With similar arguments, we decided to keep fixed also the age and metallicity values as extracted from chip 5 in the halo regions, when later performing the full-mode fit. The retrieved values of age and [Z/H], indeed, are in good agreement with the indications of the two spectral indices H$\beta$ and [MgFe]', which consolidates our choice.

To summarize, when performing the fit with ALF in full mode, as described in Section \ref{sec:alfsetup}, in the outer regions we fixed the kinematic values to the simple mode ones extracted from chip 5, while in the halos 
we fixed not only the kinematics, but also age and metallicity.

\subsection{The IMF in the centers}  
\label{app:imf}

\begin{figure}[ht!]
\begin{centering}
\includegraphics[width=10cm]{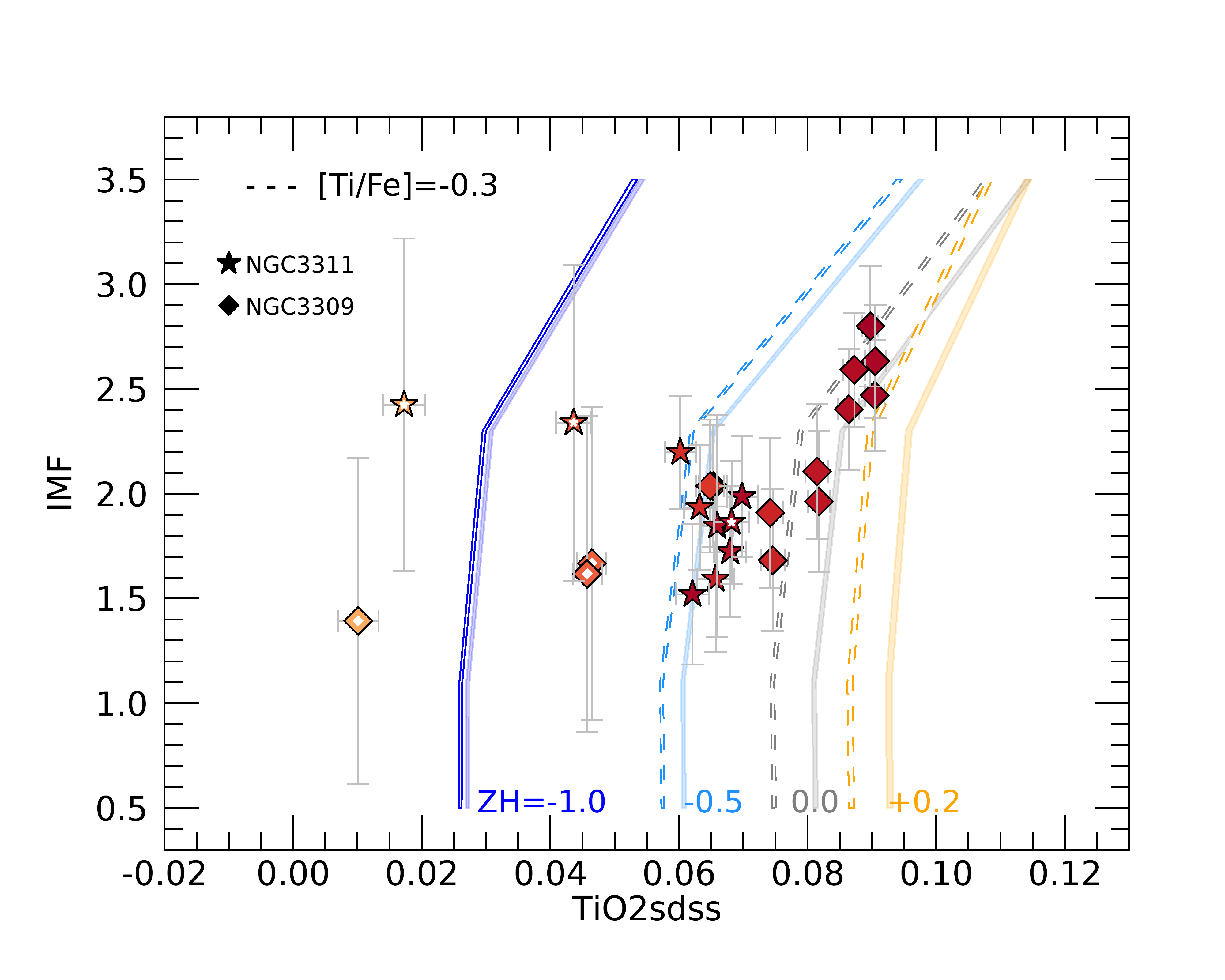}
\caption{\small{Retrieved IMF vs TiO2$_{sdss}$ index values in the centers of the two galaxies NGC3309 (diamonds) and NGC3311 (stars). Solid colored lines are the model values at different metallicity and age $13.5$ Gyr. Dashed lines show the same trends when the [Ti/Fe] abundance is sub-solar, i.e. $-0.3$ dex. All other abundances are solar. Points are color-coded as in Figure \ref{fig:ageZimf}, with darker red in the center and orange in the outskirts. While in the very central regions TiO2$_{sdss}$ values indicate a clear gradient for IMF and [Z/H] for NGC3309 and a flat trend for NGC3311 (confirmed by the retrieved results), in the outer regions where the metallicity is lower, it is harder to constrain the IMF since the same value of TiO2$_{sdss}$ corresponds to a wide range of IMF.}}
\label{fig:tio2}
\end{centering}
\end{figure}

In this section we test the accuracy of our results by comparing them with the expectations of some IMF sensitive indices, such as TiO2 and bTiO. Indeed, when taking into account age, metallicity and all of the elemental abundances (see \citealt{lonoce21}), spectral indices can give reliable indications on the IMF slope value, as widely done in the literature (e.g.: \citealt{martin15b,labarbera15,parikh18}). We measured the value of the indices in our wavelength range and made comparisons with the results obtained with ALF. In addition to  metallicity that perfectly matches the measured total metallicity indicator [MgFe]' \citep{thomas03} at the retrieved ages, also the IMF trends in the centers are confirmed, for example, by TiO2 and bTiO. In Figure \ref{fig:tio2}, the plot of the retrieved IMF slopes vs the measured TiO2$_{sdss}$ index is shown. In this plot we only show the measures from spectra that were not disturbed by sky residuals, i.e. only the central regions. Together with our measurements, we also show the expectations of models at different metallicity values (different colors) and with a depletion of [Ti/Fe] (dashed lines). We recall, however, that [Ti/Fe] is confined to a region of $\pm0.1$ dex in the center of both galaxies. The models also outline regions of different velocity dispersion values, but  notably, $\sigma$ does not affect this index much (while other indices are more affected, like bTiO).  
From this figure it is clear that the very central region of NGC3309 (dark red diamonds) has increasing values of TiO2$_{sdss}$, suggesting a degeneracy effect of both increasing [Z/H] and IMF, while NGC3311 values (dark red stars) are confined in a narrow region pointing to constant metallicity and IMF. Since [Z/H] is very well constrained as confirmed by [MgFe]', we are confident that the retrieved trend of IMF in both galaxies is well validated by the value of this index. Finally, from this plot it is also possible to understand how difficult the retrieval of the IMF at lower metallicity values is, where the slope of models is steep, meaning that the same value of TiO2$_{sdss}$ can be explained with both a Kroupa-like or bottom-heavy IMF. This is the case for the orange points, plotted with open symbols to indicate the higher uncertainty of their retrieved values.  

In conclusion, our IMF measurements in the centers of NGC3309 and NGC3311 have the following strengths: i) we fitted a large wavelength range in the optical and NIR with high S/N, including many IMF-sensitive spectral features, ii) all fits are well-converged in all parameters, including the IMF, iii) we took into account all elemental abundances that contribute to the shaping of spectral features, iv) we considered the systematics affecting all parameters, including the IMF, and v) we checked the values of IMF-sensitive spectral indices to confirm the results obtained from full spectral fitting.  We will discuss the implications of the retrieved values in Section \ref{sec:imfcor}.

\section{Fit correlations}
\label{app:fitcorr}

As mentioned in the main text, before discussing correlations among parameters, it is important to check the level of degeneracy among them during the fit, to rule out 
nonphysical correlations. In this brief section we list out results.

For each analyzed spectrum, we inspected the marginal posterior distributions of all the pairs of parameters, and to be quantitative, we calculated the Spearman correlation coefficient $\rho$ for each distribution. 

In general, each spectrum presents many relevant correlations (i.e. those with p-value $p<0.1$), but not all are recurrent for all spectra. When averaging correlations on the spectra with higher S/N (in the central regions), indeed, we found that around $30$\% of parameter pairs have a correlation with $\rho>0.20$ with p-value $p<0.1$, and only around $7$\% with $\rho>0.40$. In five cases $\rho>0.60$, they are: C-O ($\rho=0.74$, mostly due to the extraction of these two elements from the same molecule CO, see \citealt{worthey14}), Mg-Fe, Ca-Fe, Fe-[Z/H] and Na-[Z/H]. In more detail, age correlates with [Z/H] ($\rho=-0.45$) and IMF slope ($\rho=-0.48$); [Z/H] correlates mostly with Fe, Na, C, Ca, Cr, Mg, Mn, O, Si; Na correlates with [Z/H], Fe, C, Mg and O; Mg correlates also with C, Ca, Ti, V and Cr. The IMF slope correlates moderately only with age, and mildly with Na and Ti ($\rho\sim0.30$). The IMF with [Z/H] has $\rho=0.21$ but with $p=0.13$. These degeneracy indicators are similar to their analogs in \citet{barbosa21a}, obtained with different codes; however, the strength of each correlation show some differences. An example, as noted in the main text, is the correlation between [Z/H] and IMF slope which we measured $\rho=0.21$ with $p=0.15$ while \citet{barbosa21a} report $\rho=0.35$ with $p=0.00$, indicating a more important internal degeneracy in the latter.

\section{General correlations}
\label{app:gencorr}

\begin{figure*}[ht]
\begin{centering}
\includegraphics[width=18.cm]{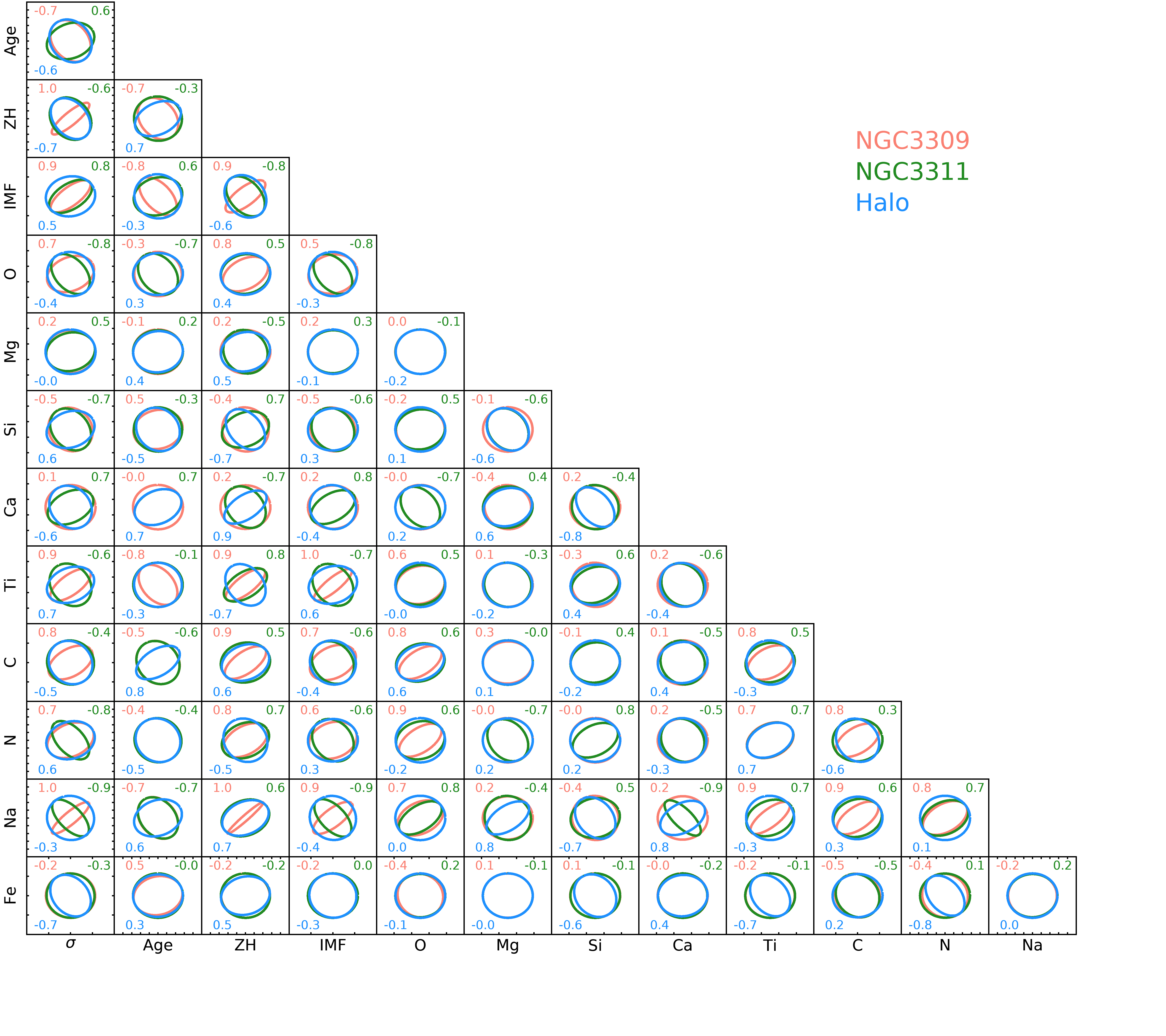}
\caption{\small{Correlations between retrieved parameters. Comparison of results obtained in the center of NGC3309 (pink), in the center of NGC3311 (green) and in the outer regions and halos (blue). Numbers refer to the Spearman correlation coefficient with the same color coding. The eccentricity of each ellipse is proportional to the calculated Spearman coefficient value. Strong correlations observed for NGC3309's center are generally not confirmed for the center of NGC3311.}}
\label{fig:cornercorr}
\end{centering}
\end{figure*}

In Figure \ref{fig:cornercorr} we show some examples of the mutual correlations between the retrieved parameters in the form of a corner plot. Ellipses indicate the degree of correlation with their ellipticity proportional to the measured Spearman rank coefficient. We divided the results in center of NGC3309 (pink), center of NGC3311 (green) and the remaining outer regions and halo (blue) to highlight their differences. In general, it can be observed that there are no strong correlations apart from the center of NGC3309. This galaxy shows strong correlations, with $\rho>0.6$, between [Z/H] and $\sigma$, Na, IMF slope, C, Ti, and O; between $\sigma$ and IMF slope, O, Ti, C, N  and Na; and as a consequence, all the combinations of these quantities. There is also an important anti-correlation between age and [Z/H], IMF slope, and $\sigma$. 

However, most of these correlations are not found in the center of NGC3311, which in some cases shows even opposite correlations. There is agreement between the two galaxies only for $\sigma$-IMF, [Z/H]-Ti, Ti-O, [Z/H]-N, Ti-N, Na-O and Na-N. On one hand, this evident difference is a sign that the kinematics radial profile plays a role in the distribution of elements. Indeed, as it will be further discussed in Section \ref{sec:scalingrelation}, there is large disagreement between the two galaxies correlations with $\sigma$, with the exception of the IMF.

On the other hand, looking more closely, for example, at correlations with the total metallicity, it can be noticed that, although they are different in the two galaxies for their slope and strength, there is a common positive trend with O, C, Ti, N and Na. This suggests that, although the metallicity trends are slightly different for the two objects, these elements track [Z/H].
This finding is in agreement with the MaNGA data results in \citet{parikh19}, in particular for Na, N and partially for Ti. However, they do not find a local correlation with C. That N follows [Z/H] is expected since N can be enhanced by a delayed secondary production, activated only at higher metallicity (\citealt{johansson12,maiolino19}, and references therein).

In the halo regions we do not see many strong correlations, with the exception of e.g.: Ca-[Z/H], Ca-Si, C-age, Na-Mg, Na-Ca, Fe-N; we only note the excellent accord among all regions in the correlation between N and Ti. This general disagreement between central and outer regions indicates likely different origins and star formation histories of their stellar content.  

Comparing the correlations in the centers of the two galaxies with our results described in \citet{feldmeier21} obtained from a sample of local ellipticals, we confirm the correlations between Na-O, Na-N, C-O (but affected by cross-correlation in the fit),  N-O, Na-O, C-Ti, Na-C. We also observe the same [Z/H]-age anti-correlation, but much of it comes from the fit degeneracy.

\bibliography{Hydra_biblio}{}
\bibliographystyle{aasjournal}



\end{document}